\documentclass[3p]{elsarticle}

\usepackage{amssymb,bm}
\usepackage{subfigure}
\usepackage{subfigmat}
\usepackage{graphicx}
\usepackage{color}

\newcommand{\sgn}{\mathop{\mathrm{sgn}}}
\newcommand{\rd}{\mathrm{d}}

\newcommand{\LLL}{\\ \newline}
\newcommand*{\revA}{}

\journal{Journal of Fluids and Structures}

\begin{document}
\begin{frontmatter}
\title{Two-domain and three-domain limit cycles in a typical aeroelastic system with freeplay in pitch} 
 \author[edref]{Edouard Verstraelen}
\author[gregref]{Grigorios Dimitriadis\corref{correspondingauthor}}
\author[gustavoref]{Gustavo Dal Ben Rossetto}
\author[earlref]{Earl H. Dowell}

\cortext[correspondingauthor]{Corresponding author, gdimitriadis@ulg.ac.be}
\fntext[edref]{PhD Candidate, University of Li\`ege, All\'ee de la D\'ecouverte 9, Belgium}
\fntext[gregref]{Associate Professor, University of Li\`ege, All\'ee de la D\'ecouverte 9, Belgium, Senior Member AIAA.}
\fntext[gustavoref]{Aeronautics Engineer, Technology Development Department, Av. Brigadeiro Faria Lima, 2170, S\~ao Jos\'e dos Campos, S\~ao Paulo, 12227-901, Brazil}
\fntext[earlref]{William Holland Hall Professor, Duke University, Department of Mechanical Engineering and Materials Science, Box 90300, Hudson Hall, Durham, NC, USA, Grade Honorary Fellow AIAA}


\begin{abstract}
Freeplay is a significant source of nonlinearity in aeroelastic systems and is strictly regulated by airworthiness authorities. It splits the phase plane of such systems into three piecewise linear subdomains. Depending on the location of the freeplay, limit cycle oscillations can result that span either two or three of these subdomains. The purpose of this work is to demonstrate the existence of two-domain cycles both theoretically and experimentally. A simple aeroelastic system with pitch, plunge and control deflection degrees of freedom is investigated in the presence of freeplay in pitch. It is shown that two-domain and three-domain cycles can result from a grazing bifurcation and propagate in the decreasing airspeed direction. Close to the bifurcation, the two limit cycle branches interact with each other and aperiodic oscillations ensue. Equivalent linearization is used to derive the conditions of existence of each type of limit cycle and to predict their amplitudes and frequencies. Comparisons with measurements from wind tunnel experiments demonstrate that the theory describes these phenomena with accuracy.
\end{abstract}


\end{frontmatter}


\section{Introduction}

Freeplay in actuators and bearings is a significant source of nonlinearity in aeroelastic systems. Airworthiness authorities place very strict limits on the amount of freeplay allowed in aircraft control surfaces (see~\cite{FAA_ANM_019} for example). Assessing the impact of freeplay on aeroelastic responses is therefore an important aspect of nonlinear aeroelastic research.
\LLL
\revA{Numerous experimental and numerical studies have been published in the literature on the aeroelastic behaviour of simple systems with freeplay nonlinearity, paying particular attention to Limit Cycle Oscillation (LCO) phenomena. The effect of freeplay was investigated on rigid wings with pitch and plunge degrees-of-freedom (DOFs) with freeplay in the pitch DOF~\cite{yang1,hauenstein,price1,Lee3,Lee2,marsden1,chung1,vasconcellos} but also in the plunge DOF~\cite{hauenstein} or with an external store added~\cite{liu1}. Tang et al. also investigated a similar system with a flap instead of a plunge DOF~\cite{tang3}. A flexible aerodynamic surface attached to a pitch shaft with freeplay was also considered~\cite{chen1,tang2,tang1} and Lee et al. even added a plunge DOF~\cite{Lee1}. Another well known aeroelastic apparatus is the typical aeroelastic section (a rigid wing with pitch, plunge and control DOFs) with freeplay in the control surface~\cite{Conner97,alighanbari,gordon1,manetti1,vasconcellos5,kholodar,cui1,pereira1}. Most of these systems were shown to undergo several different \revA{types} of LCO, including asymmetric and aperiodic oscillations at airspeeds lower than their linear flutter speed. Nevertheless, all the LCOs orbit the single fixed point lying at the origin.}
\LLL
In this paper, \revA{the typical aeroelastic section} is investigated but with freeplay in the pitch DOF instead of the control surface. It is shown that \revA{this location} of the freeplay can cause anti-symmetric fixed points to come into existence and that limit cycles can \revA{orbit} these fixed points. This phenomenon is interesting because the system is completely symmetric and yet gives rise to asymmetric periodic responses and fixed points. The emergence of symmetric and asymmetric limit cycles is investigated for both symmetric freeplay and freeplay with preload. The phenomena are demonstrated mathematically and experimentally by means of an aeroelastic wind tunnel model.

\section{General equations of motion of an aeroelastic system with freeplay and preload} 

\begin{figure}[ht]
  \begin{center}
    \includegraphics[width=.7\textwidth]{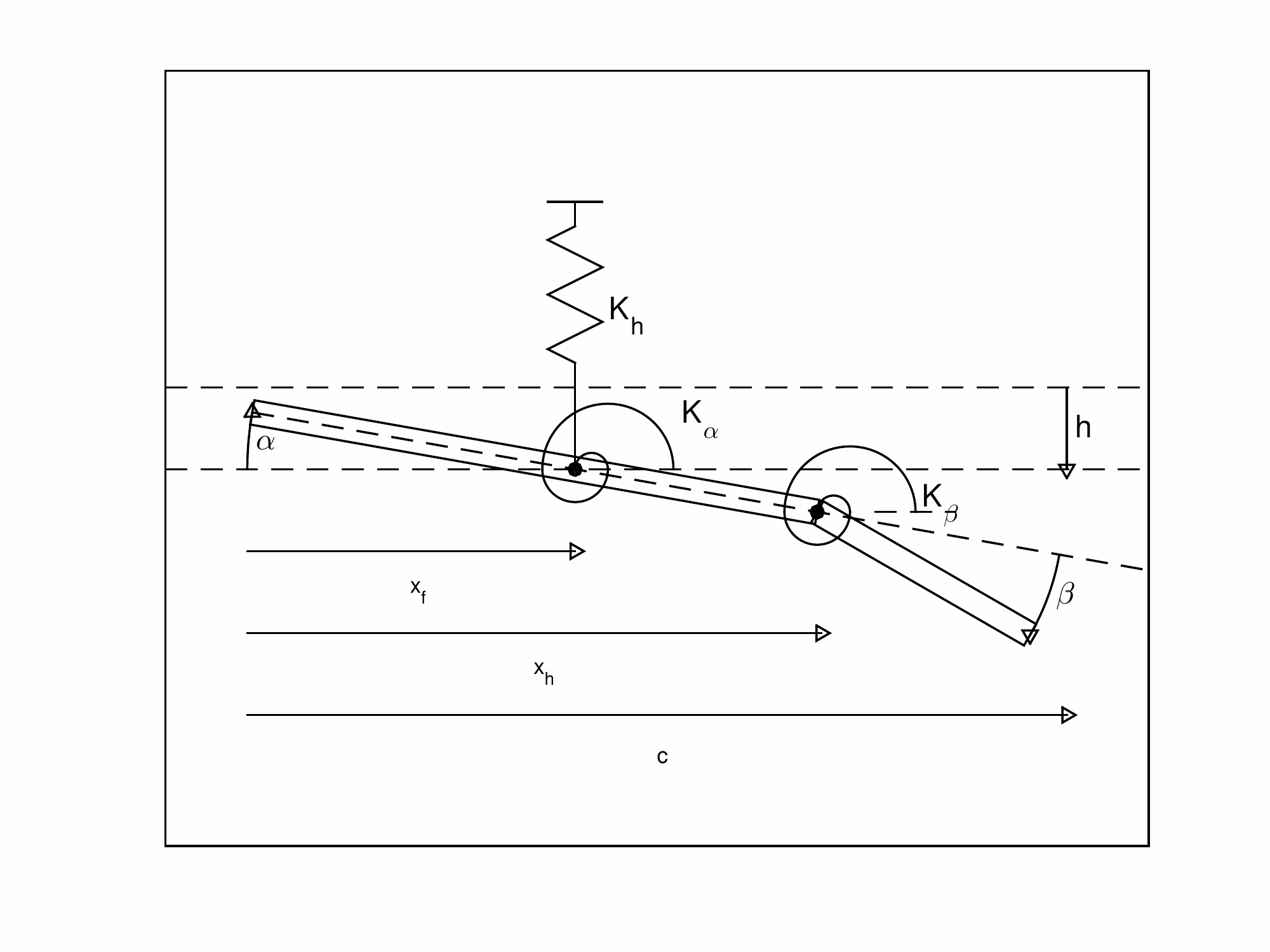}
  \end{center}
  \caption{Pitch-plunge-control aeroelastic system}
\label{fig_dof3}
\end{figure}

\begin{figure}[ht]
  \begin{center}
    \includegraphics[width=.47\textwidth]{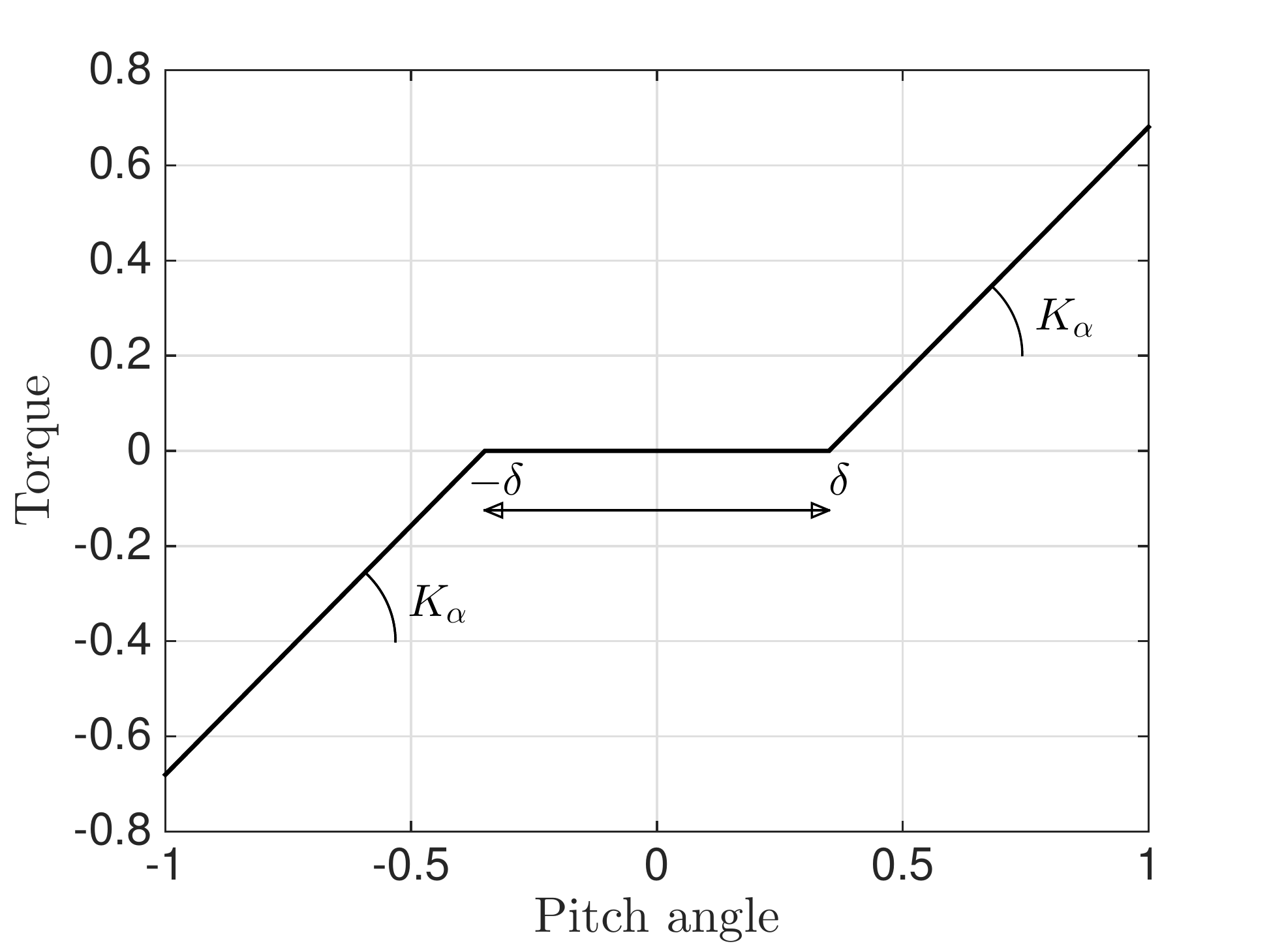}
  \end{center}
  \caption{Freeplay stiffness diagram}
\label{fig_freeplay}
\end{figure}

The pitch-plunge-control aeroelastic system is a 2D symmetric flat plate wing with a control surface. The entire wing is suspended by an extension spring with stiffness $K_h$ and a rotational spring of stiffness $K_{\alpha}$ from its pitch axis $x_f$. These two springs provide restoring forces in the plunge, $h$, and pitch, $\alpha$, DOFs respectively. The control surface deflection angle $\beta$ is an additional DOF, restrained by a rotational spring with stiffness $K_{\beta}$. The control surface hinge lies at $x_h$ and the total chord of the wing is denoted by $c$. The complete system is shown in figure~\ref{fig_dof3}
\LLL
It is assumed that there is freeplay in the pitch DOF, such that the \revA{restoring force in the corresponding spring} is zero while $|\alpha| < \delta$, $2\delta$ being the width of the freeplay region. Figure~\ref{fig_freeplay} shows a typical restoring force diagram for freeplay, whereby the stiffness is $K$ \revA{if} $|\alpha| > \delta$ and zero otherwise. Note that the freeplay region is centred around the origin.
\LLL
In the case of the pitch-plunge-control wing with freeplay in the pitch DOF, the stiffness outside the freeplay region is given by $K_{\alpha}$, while the stiffness inside the freeplay region is zero. The restoring moment equation is
\begin{equation}
M_{\alpha}(\alpha)=\left\{
\begin{array}{cc}
K_{\alpha}(\alpha+\delta) & \mbox{if \hspace{2ex}} \alpha < -\delta \\
0 & \mbox{if \hspace{2ex}} |\alpha| \leq \delta \\
K_{\alpha}(\alpha-\delta) & \mbox{if \hspace{2ex}} \alpha > \delta
\end{array}
\right.
\label{eq_freeplay}
\end{equation}
where $M_{\alpha}$ is the pitching moment provided by the freeplay spring. 
%
%
\LLL
\revA{In addition to the freeplay, an aerodynamic preload angle $\alpha_p$ is considered. It models the fact that when the wing is perfectly centred in the freeplay region, it is not perfectly aligned with the airflow and vice versa. As a result, the structural moment, $M_\alpha$, depends on $\alpha(t)$ while the aerodynamic moment, $M_{aero}$, depends on $\alpha_{tot} = \alpha_p + \alpha(t)$, which introduces an asymmetry in the system.}
\LLL
The equations of motion of the system flying with airspeed $U$ in air of density $\rho$ can be developed using linear unsteady attached flow aerodynamic assumptions; a time-domain model can be written by means of Wagner function analysis~\cite{Conner97,Dimitriadis11}. The structural displacements are denoted by the vector $\mathbf{y}=[ h \mbox{\hspace{2ex}} \alpha \mbox{\hspace{2ex}}  \beta ]$ while the \revA{six} aerodynamic states are denoted by the vector $\mathbf{ w}=[ w_1 \mbox{\hspace{2ex}} \ldots \mbox{\hspace{2ex}}  w_6 ]$. Then the complete state vector of the system is given by $\mathbf{x}=[\dot{\mathbf{y}} \mbox{\hspace{2ex}} \mathbf{y} \mbox{\hspace{2ex}} \mathbf{ w}]^T$ and has dimensions $12\times 1$. The equations of motion of the system with freeplay and aerodynamic preload in the pitch DOF are given by
\begin{equation}
\dot{\mathbf{x}}=
\mathbf{Q}_1\mathbf{x} + \mathbf{q}_n M_{\alpha}(\alpha)  +\mathbf{q}_p \alpha_p
\label{eq_dof3_freeplay_alpha}
\end{equation}
where
\begin{eqnarray} 
 \mathbf{Q}_1 & = & \left(
\begin{array}{ccc}
-\mathbf{M}^{-1}\left(\mathbf{C}+\rho U \mathbf{D} \right) & -\mathbf{M}^{-1}\left(\mathbf{E}_{1}+\rho U^2 \mathbf{F} \right) & -\rho U^3 \mathbf{M}^{-1} \mathbf{W} \\
\mathbf{I}_{3\times 3} & \mathbf{0}_{3\times 3} & \mathbf{0}_{3\times 6} \\
\mathbf{0}_{6\times 3} & \mathbf{W}_1 & U\mathbf{W}_2
\end{array}
\right) 
\nonumber\\
\mathbf{q}_n & = &
\left(
\begin{array}{c}
-\mathbf{M}^{-1}
\left(
\begin{array}{c}
0\\
1\\
0
\end{array}
\right) \\
\mathbf{0}_{9\times 1}
\end{array}
\right) \nonumber\\
\mathbf{q}_p & = &
\left(
\begin{array}{c}
\rho U^2 \mathbf{M}^{-1} \bm{P}  
 \\
\mathbf{0}_{9\times 1}
\end{array}
\right)
\label{eq_dof3_qnonlin_freepl}
\end{eqnarray} %
and $\mathbf{E}_{1}$, is the value of the structural stiffness matrix inside the freeplay region $\pm\delta$, given by
\begin{equation}
 \mathbf{E}_1  =  \left(
\begin{array}{ccc}
K_h & 0 & 0 \\
0 & 0 & 0 \\
0 & 0 & K_{\beta}
\end{array}
\right)  
\end{equation}
\LLL
Matrix $\mathbf{C}$ is the structural damping matrix, $\rho U \mathbf{D}$ is the aerodynamic damping matrix, $\rho U^2 \mathbf{F}$ is the aerodynamic stiffness matrix, $\mathbf{W}$ is the aerodynamic state matrix, $\mathbf{W}_1$ and $\mathbf{W}_2$ are the aerodynamic state equation matrices, $\mathbf{M}=\mathbf{ A}+\rho \mathbf{ B}$, $\mathbf{ A}$ is the structural mass matrix and $\mathbf{B}$ is the aerodynamic mass matrix. The matrix $\mathbf{P}$ is an aerodynamic stiffness vector that takes into account the \revA{effect} of the preload angle $\alpha_p$ on the loads acting on the system. The notation $\mathbf{I}_{3\times 3}$ denotes a unit matrix of size $3\times 3$. The values of all the matrices are given in Appendix A. \revA{Equation}~\ref{eq_dof3_freeplay_alpha} can be written as

\revA{
\begin{equation}
\dot{\mathbf{x}} = \left \{
\begin{array}{llc}
\mathbf{Q}_1\mathbf{x}+\mathbf{q}_p \alpha_p & \mbox{\hspace{2ex}  if \hspace{2ex}} |\alpha| \leq \delta & \mbox{\hspace{2ex} (a)}\\
\mathbf{Q}_2\mathbf{x}-\mathbf{q}_n K_{\alpha} \sgn(\alpha) \delta+\mathbf{q}_p \alpha_p & \mbox{\hspace{2ex} if \hspace{2ex}} |\alpha| > \delta & \mbox{\hspace{2ex} (b)}
\end{array}
\right.
\label{eq_dof3_23}
\end{equation}}
%

\noindent
\revA{where $\mathbf{Q}_2 \mathbf{x}=\mathbf{Q}_1 \mathbf{x} + \mathbf{q}_n K_{\alpha} \alpha$. 
\LLL
In this work, we will define two linear sub-systems that are relevant to freeplay:
\begin{itemize}
\item Underlying linear system: the system without structural stiffness that is only valid inside the freeplay region (equation \ref{eq_dof3_23}(a)). 
\item Overlying linear system: the nominal system without freeplay and with full stiffness (equation \ref{eq_dof3_23}(b) with $\delta$ = 0).
\end{itemize}}

\clearpage
\section{Fixed points} 

The freeplay function of figure~\ref{fig_freeplay} splits the phase plane of the system responses into three piecewise linear subdomains, $S_1$ for $|\alpha| \leq \delta$, $S_2$ for $\alpha > \delta$ and $S_3$ for $\alpha < -\delta$. Response trajectories can span one, two or all three of the subdomains. Furthermore, equation~\ref{eq_dof3_freeplay_alpha} has three fixed points given by
\begin{eqnarray}
\mathbf{x}_{F_1} & = &   -\mathbf{Q}_1^{-1}\mathbf{q}_p \alpha_p  \mbox{\hspace{2ex} if \hspace{2ex}} |\alpha| \leq \delta \nonumber \\
\mathbf{x}_{F_2} & = &   \mathbf{Q}_2^{-1} \left( \mathbf{q}_n K_{\alpha}\delta -\mathbf{q}_p \alpha_p \right)  \mbox{\hspace{2ex} if \hspace{2ex}} \alpha > \delta \label{eq_xf} \label{eqn_xf_1}\\
\mathbf{x}_{F_3} & = &   -\mathbf{Q}_2^{-1}  \left( \mathbf{q}_n K_{\alpha}\delta +\mathbf{q}_p \alpha_p \right)  \mbox{\hspace{2ex} if \hspace{2ex}} \alpha < -\delta \nonumber 
\end{eqnarray}
i.e. they depend on the aerodynamic preload $\alpha_p$, the freeplay gap $\delta$ and the airspeed $U$. These fixed points do not coexist; only one of them is an attractor at any instance in time, depending on which subdomain the response trajectory lies in. \revA{Dividing equations \ref{eqn_xf_1} throughout by $\delta$} we obtain \revA{non-dimensional} fixed points $\bar{\mathbf{x}}_{F_i}=\mathbf{x}_{F_i}/\delta$ that only depend on the airspeed and the ratio of the aerodynamic preload divided by the freeplay gap, $\alpha_p/\delta$.
\begin{figure}[ht]
  \begin{center}
    \subfigure[Fixed points]{\label{sfig_dof3_freepl_alpha_xf} \includegraphics[width=.47\textwidth]{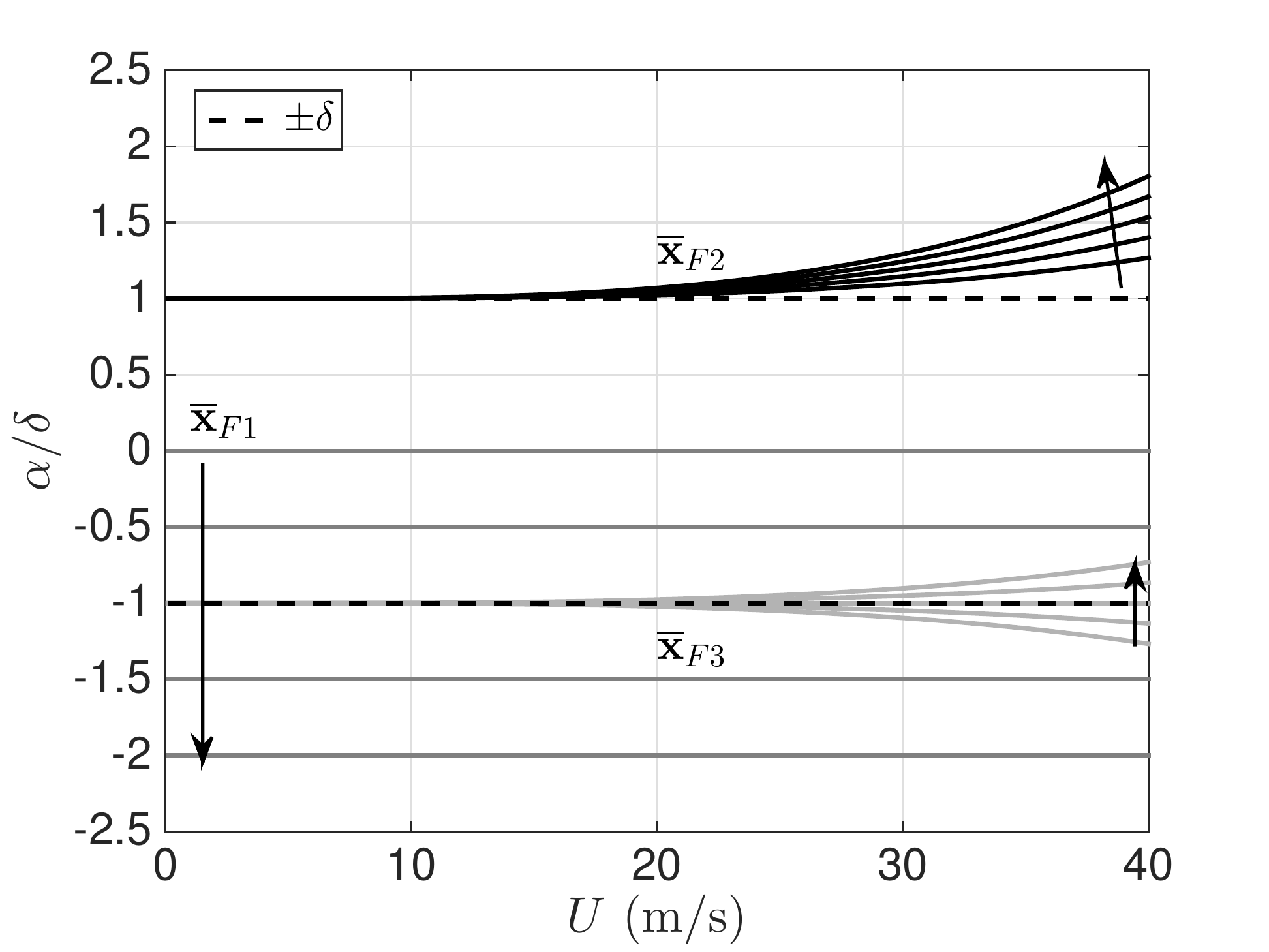}}
    \subfigure[Boundary-equilibrium bifurcation]{\label{sfig_subdomains} \includegraphics[width=.47\textwidth]{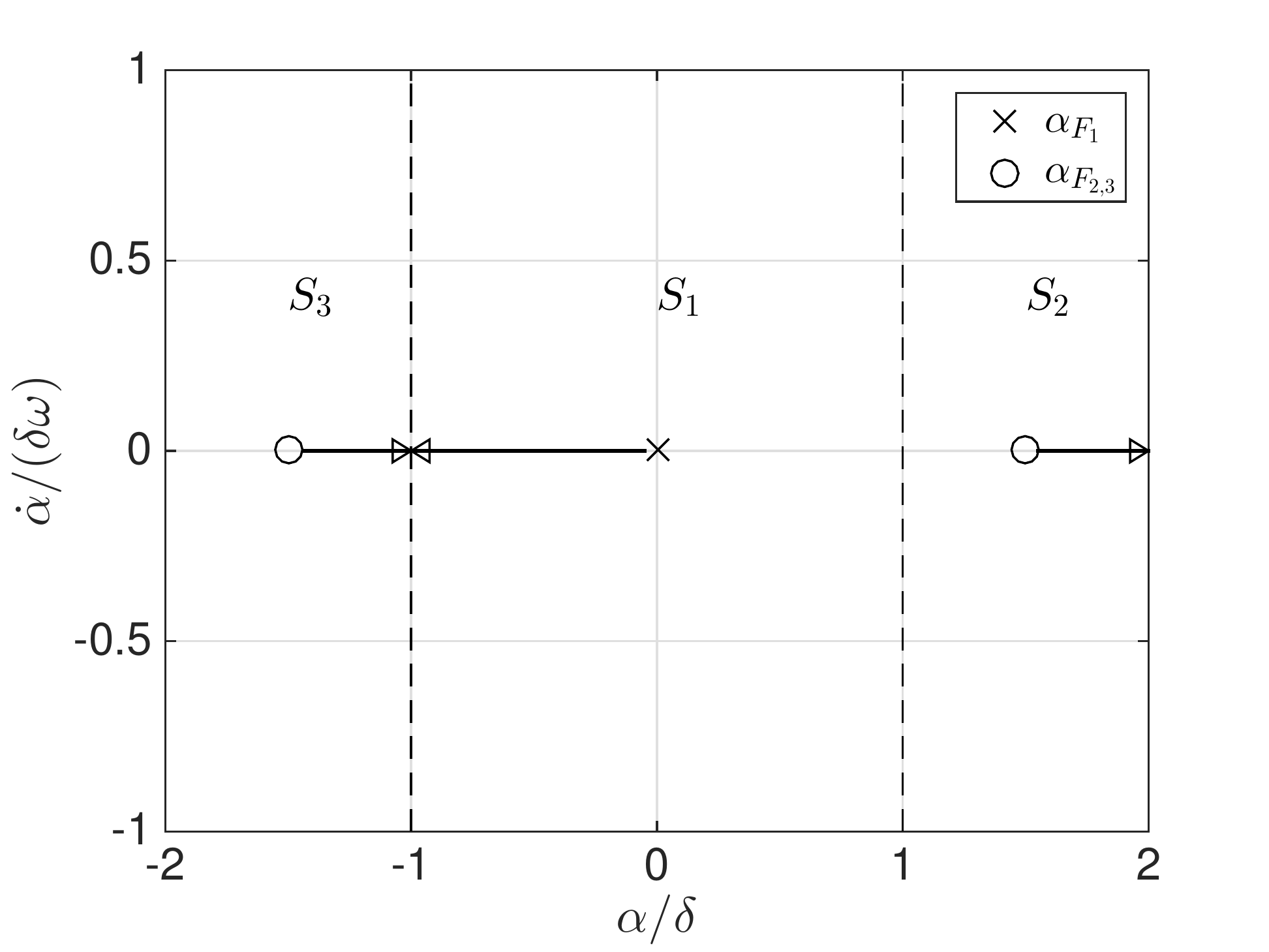}}
  \end{center}
  \caption{Positions of fixed points with varying airspeed and $\alpha_p/\delta$ ratio}
\label{fig_subdomains}
\end{figure}
\LLL
Figure~\ref{sfig_dof3_freepl_alpha_xf} plots the pitch component of the three fixed points for airspeeds between 0 and 40 m/s and for $\alpha_p/\delta$ ratios between 0 and 2. It can be seen that, as the aerodynamic preload ratio increases, fixed point $\mathbf{\overline{x}}_{F_1}$ moves from 0 to -1 and eventually exits the freeplay region. Similarly, $\mathbf{\overline{x}}_{F_3}=1$ for all airspeeds when  $\alpha_p/\delta=1$ and enters the freeplay region for all $\alpha_p/\delta>1$. This means that only $\mathbf{\overline{x}}_{F_2}$ exists for $\alpha_p/\delta>1$, since  $\mathbf{\overline{x}}_{F_1}$ and $\mathbf{\overline{x}}_{F_3}$ violate the conditions for existence given in equations~\ref{eq_xf}.
\LLL
\revA{As the aerodynamic preload increases,} the system bifurcates from a system with three fixed points to a system with 1 fixed point. This bifurcation is known as a boundary-equilibrium bifurcation and occurs when $\alpha_p=\delta$ and $\mathbf{x}_{F_1}=\mathbf{x}_{F_3}=-\delta$ for all airspeeds. The bifurcation can be more easily visualised in the phase plane plot of figure~\ref{sfig_subdomains}. The system's fixed points are plotted in the $\alpha$-$\dot{\alpha}$ plane for $\alpha_p=0$ and \revA{$U\neq 0$}. The arrows denote the motion of the fixed points as $\alpha_p$ is increased. \revA{As mentioned earlier, the freeplay} region divides the phase plane into three subdomains: $S_1$ inside the freeplay boundaries at $\pm 1$, $S_2$ and $S_3$ outside. Each fixed point is only defined inside its respective subdomain. The arrows show that, as $\alpha_p$ increases, fixed points $\mathbf{x}_{F_1}$ and $\mathbf{x}_{F_3}$ collide with the $-\delta$ boundary and disappear. Another boundary-equilibrium bifurcation occurs at $U=0$, where $\mathbf{x}_{F_2}=\delta$, $\mathbf{x}_{F_3}=-\delta$ for all values of the aerodynamic preload ratio. 
\LLL
The fixed points of systems with piecewise linear stiffness display a transient characteristic. In the present case there are three piecewise linear subdomains and three piecewise linear systems. The fixed point of each system exists and \revA{attracts} the response trajectory while the latter lies in the corresponding subdomain. There are two types of fixed point:
\begin{itemize}
\item Fixed point of system in $S_i$ that lies in subdomain $S_i$.
\item Fixed point of system in $S_i$ that lies in another subdomain.
\end{itemize}
The first type of fixed point can attract static solutions, i.e. the system response can decay towards it (or, if the response trajectory starts on the fixed point it will stay on it forever). The second type of fixed point cannot attract static solutions and therefore is not a fixed point in the classic sense. However, it can still attract dynamic solutions while the response trajectory travels through subdomain $S_i$. The \revA{present} discussion of the boundary-equilibrium bifurcation only concerns static solutions. Indeed, when a fixed point crosses a discontinuity boundary into the wrong subdomain it cannot attract static solutions anymore. It disappears in a static sense but it still affects the response intermittently every time a trajectory enters the relevant subdomain.

\clearpage
\section{Two-domain and three-domain limit cycles} 

\begin{figure}[ht]
  \begin{center}
 \subfigure[Grazing bifurcation]{\label{sym_grazing1} \includegraphics[width=.47\textwidth]{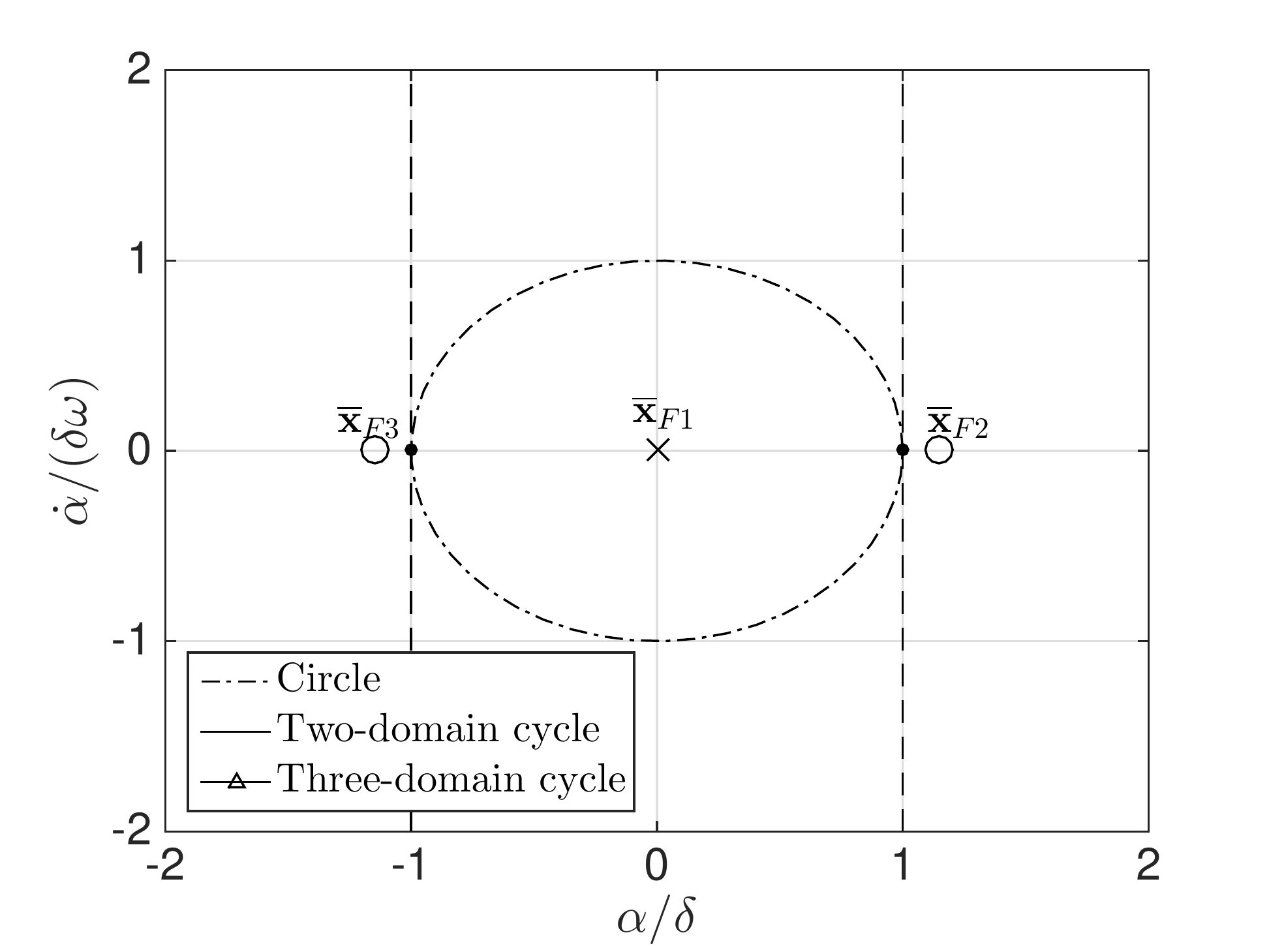}}
 \subfigure[Two-domain and Three-domain cycles]{\label{sym_grazing2} \includegraphics[width=.47\textwidth]{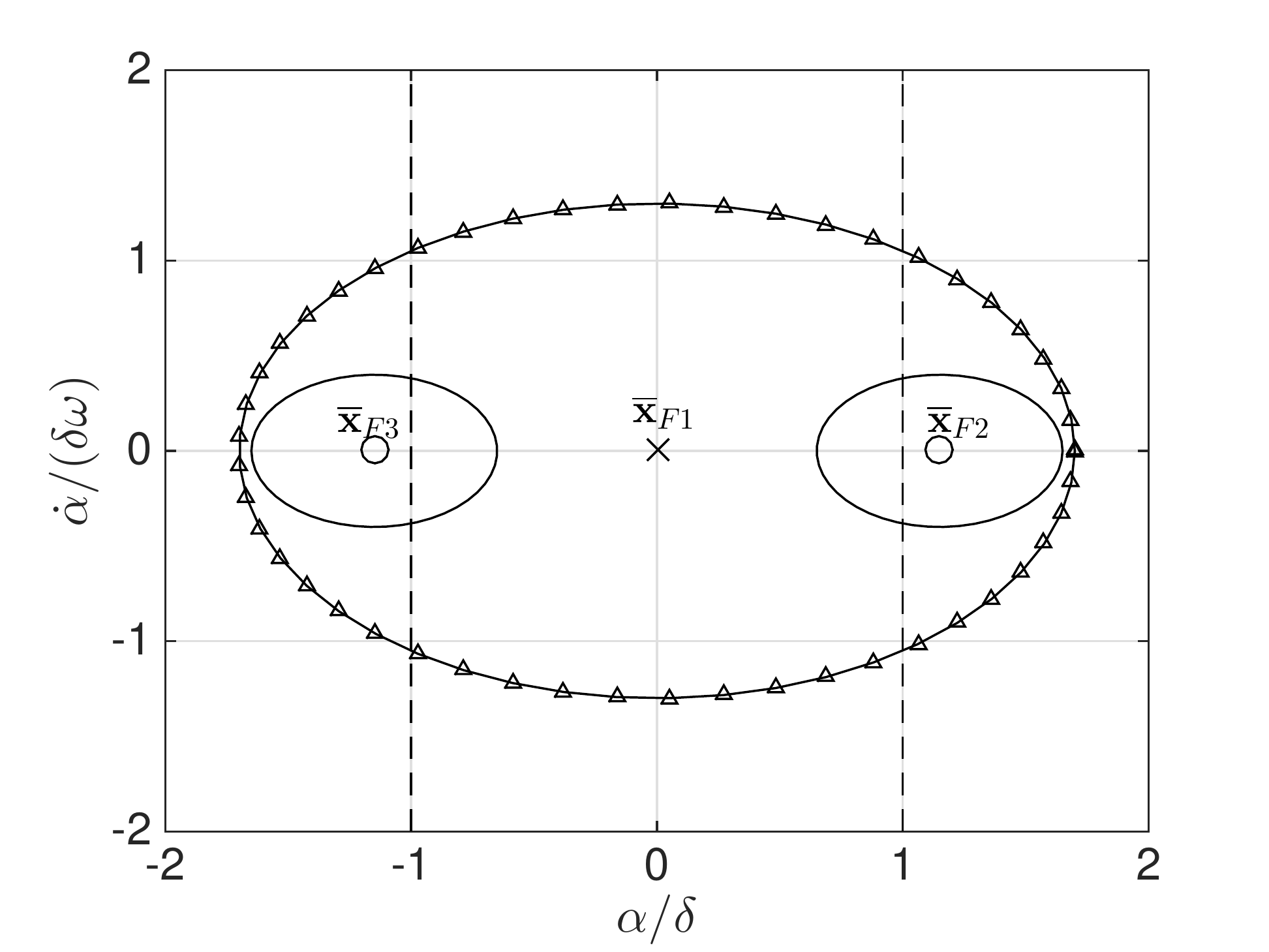}}
  \end{center}
 \caption{Grazing bifurcation fo the symmetric system}
 \label{sym_grazing}
\end{figure}

It is reasonable to make the assumption that the overlying (i.e. nominal) linear system is flutter-free and divergence-free \revA{inside its flight envelope} and that any aeroelastic instabilities are due to the freeplay. Furthermore, it will be assumed that the flutter speed of the underlying linear system $U_{F_1}$ is lower than that of the overlying linear system, $U_{F_2}$. Two types of periodic solution are then possible:
\begin{itemize}
\item Circles: These exist entirely in the $S_1$ subdomain and can only occur at the flutter point of the underlying linear system, i.e. when $U=U_{F_1}$. 
\item Limit cycles: These must span at least two subdomains as they can only exist if the system \revA{response} is nonlinear. They can exist at a range of airspeeds.
\end{itemize}
The circles and limit cycles are related; the circles bifurcate into limit cycles when their amplitude becomes equal to the width of the freeplay boundary, \revA{as displayed in figure \ref{sym_grazing1}.} This bifurcation is known as a grazing bifurcation (see for example~\cite{diBernardo08}). Limit cycles that span two domains, i.e. $S_1$ and $S_2$ or $S_1$ and $S_3$ are referred to as two-domain cycles. Limit cycles that span all three domains are referred to as three-domain cycles. Figure~\ref{sym_grazing2} plots both types of limit cycle. In the absence of aerodynamic preload, it can be seen that a three-domain cycle will orbit $\mathbf{x}_1$ and $\mathbf{x}_{2,3}$ if they exist.  In contrast, a two-domain cycle can only orbit either $\mathbf{x}_2$ or $\mathbf{x}_3$. It follows that two-domain cycles can only exist if the fixed points $\mathbf{x}_{2,3}$ also exist. 

\clearpage
\section{Experimental setup} 

\begin{figure}[ht]
  \begin{center}
  \subfigure[Wing in the test section of the wind tunnel]{\label{fig_wing} \includegraphics[height=.4\textwidth]{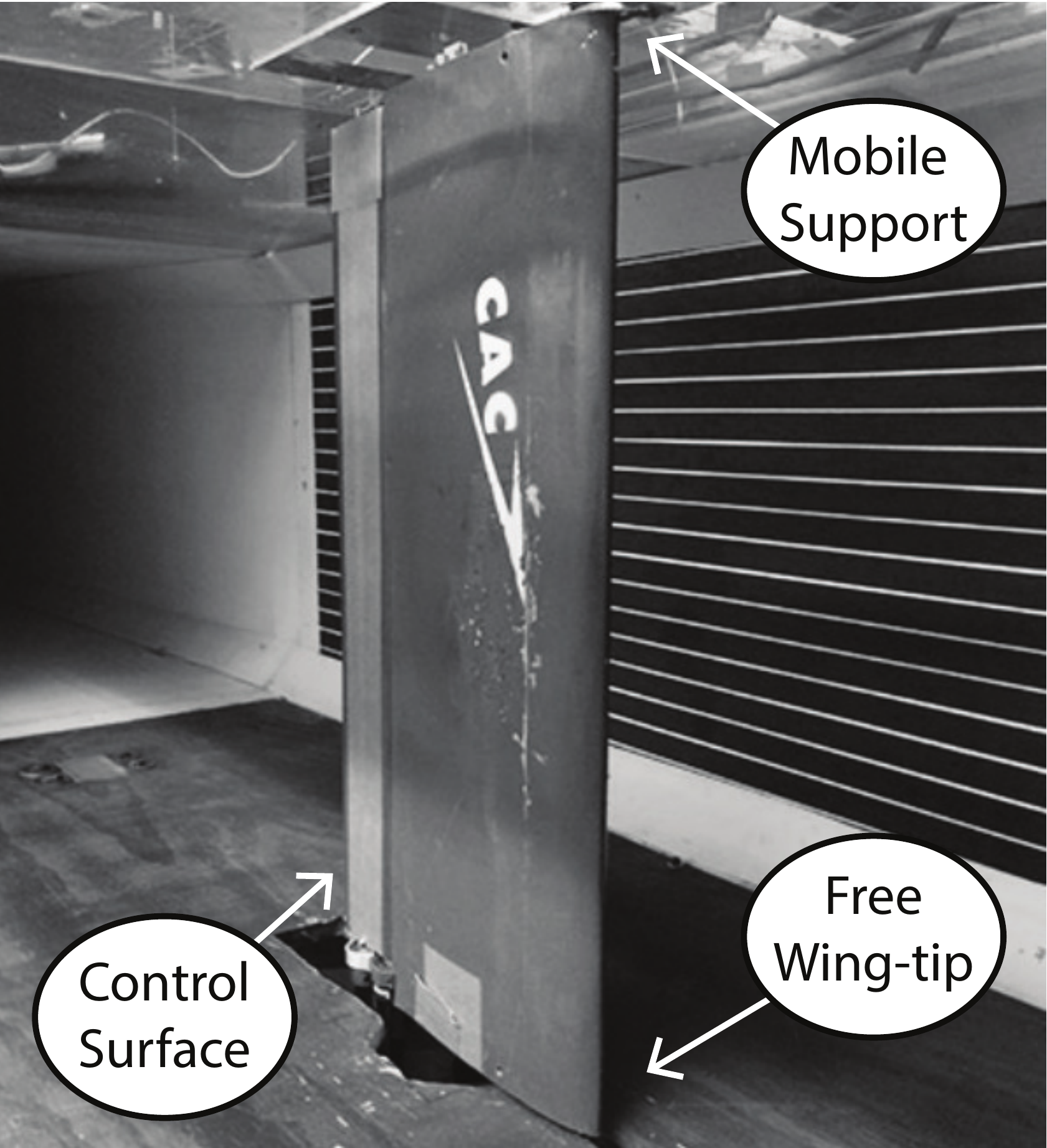}}
\subfigure[Mobile support structure]{\label{fig_mobile_sup} \includegraphics[height=.4\textwidth]{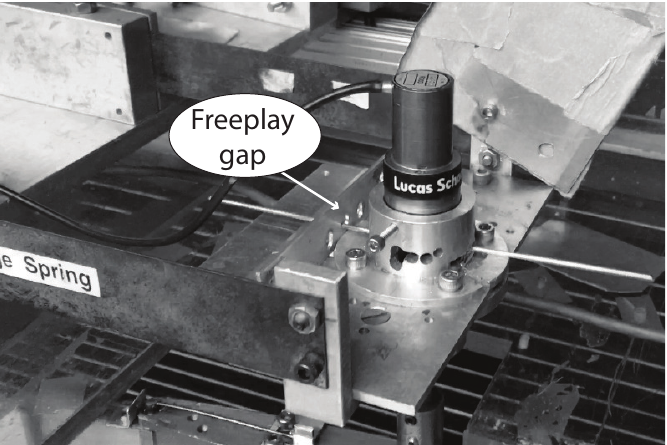}}
  \end{center}
  \caption{Photograph of the experimental apparatus}
\label{support_photo}
\end{figure}

The experimental system consists of a wing with pitch, plunge and control DOFs mounted vertically in the low-speed wind tunnel of Duke University. The two tips of the wing lie very close to the tunnel wall, so that the flow can be considered quasi-2D. For this study, the \revA{freeplay} is placed in the pitch DOF. The wing is a NACA 0012 with span 52cm and chord 19cm with a flap of chord 6.35cm mounted to the wing's trailing edge with micro-bearings and a pin. The model is externally similar to the one used by Conner et al.~\cite{Conner97}, however the support system and internal structures are different so the dynamic behaviours of the two models are different. \revA{The restoring force in the control DOF is provided by a thin piano wire glued to the pin} and clamped on the main wing. The wing is attached to the mobile support structure \revA{by means of a} single spar located at $x_f = 0.25c$ and two bearings placed on the support. Tightening screws allow the user to change the alignment of the wing in the test section to modify the aerodynamic preload angle $\alpha_p$. In this paper, angles of approximately 0 and 5 degrees are considered. Another piano wire is used to provide the necessary restoring torque in pitch. Finally, the mobile support is clamped \revA{to the wind tunnel by means of} two leaf springs that provide restoring force in the plunge DOF. Five configurations of pitch restoring forces are considered. Firstly the wire was clamped without any freeplay in order to study the overlying linear system. Then it was inserted into plates with holes of different dimensions in order to introduce different amounts of freeplay. The \revA{nominal freeplay ranges}, defined as the nominal angular distance between the two ends of the freeplay range, considered in this study are given in table \ref{table_NFR}, along with the measured freeplay ranges and the corresponding values of $\delta$. The freeplay range measurements were obtained using \revA{a pitch angular sensor}, by moving the piano wire manually between the two ends of the freeplay range. This type of measurement is prone to experimental error therefore a range of values is given in the table.
\begin{table}[h!]
\center
\begin{tabular}{ccc}
Nominal freeplay (deg) & Measured freeplay (deg) & $\delta$ (deg) \\
\hline
1 & 0.6 - 0.9 & 0.3 - 0.45 \\
2 & 1.8 - 2.0 & 0.9 - 1 \\
3 & 3.1 - 3.2 & 1.55 - 1.6 \\
8 & 7.4 - 7.6 & 3.7 - 3.8 \\
\hline
\end{tabular}
\caption{Nominal freeplay of the system} \label{table_NFR}
\end{table}
\LLL
\revA{The} structural parameters of the system\revA{, given in appendix B,}  were identified using different techniques. The inertia and stiffness of the different components were measured statically while dynamic tests were performed for validation and for damping computation. Modal analysis performed using the least square complex exponential method at several airspeeds on the overlying linear system highlighted the following modal characteristics: 
\begin{itemize}
\item The plunge-dominated mode has a frequency of 2.9~Hz and a damping of 0.87 \% at wind-off conditions. Increasing the airspeed has a hardening effect that makes it interact with the pitch mode. 
\item The pitch-dominated mode's wind-off frequency is 7.1~Hz and its damping is equal to 1.39 \%. The airflow has a softening effect on this mode that helps the interaction with the plunge mode. 
\item The flap-dominated mode lies at 17 Hz at wind-off with a damping of 0.6 \%. The airspeed has a softening effect on this mode, however its frequency is too high to allow interactions with any of the other modes in the airspeed range considered here. 
\end{itemize}
A flutter test performed on the overlying linear system showed that hard flutter occurs at 27-28~m/s due to the interaction of the pitch and plunge modes. All the experiments with freeplay are performed at airspeeds between 8 and 20~m/s\revA{, which lie far} below the flutter speed of the overlying linear system and \revA{are} therefore safe. 
\LLL
The system is instrumented in pitch using a meas-spec R30D rotation sensor with a sensitivity of 0.125~mV/deg. A meas-spec R30A angular sensor with sensitivity 0.02~mV/deg is used to measure the flap angle because of its smaller size. Finally an ultrasound sensor with a sensitivity of 10~V/m measures the plunge \revA{displacement.} The airspeed in the wind-tunnel is measured in \revA{real time} using a hot wire probe. The data from all these instruments are acquired simultaneously on a NI CompactDAQ \revA{with a sampling frequency of 1~kHz}. All the signals are 20~seconds long and \revA{low-pass} filtered at 45 Hz in order to remove electrical noise. 

\clearpage
\section{Experimental results} 

\begin{figure}[ht]
  \begin{center}
  \subfigure[Amplitude]{\label{bifA_AOA0} \includegraphics[height=.35\textwidth]{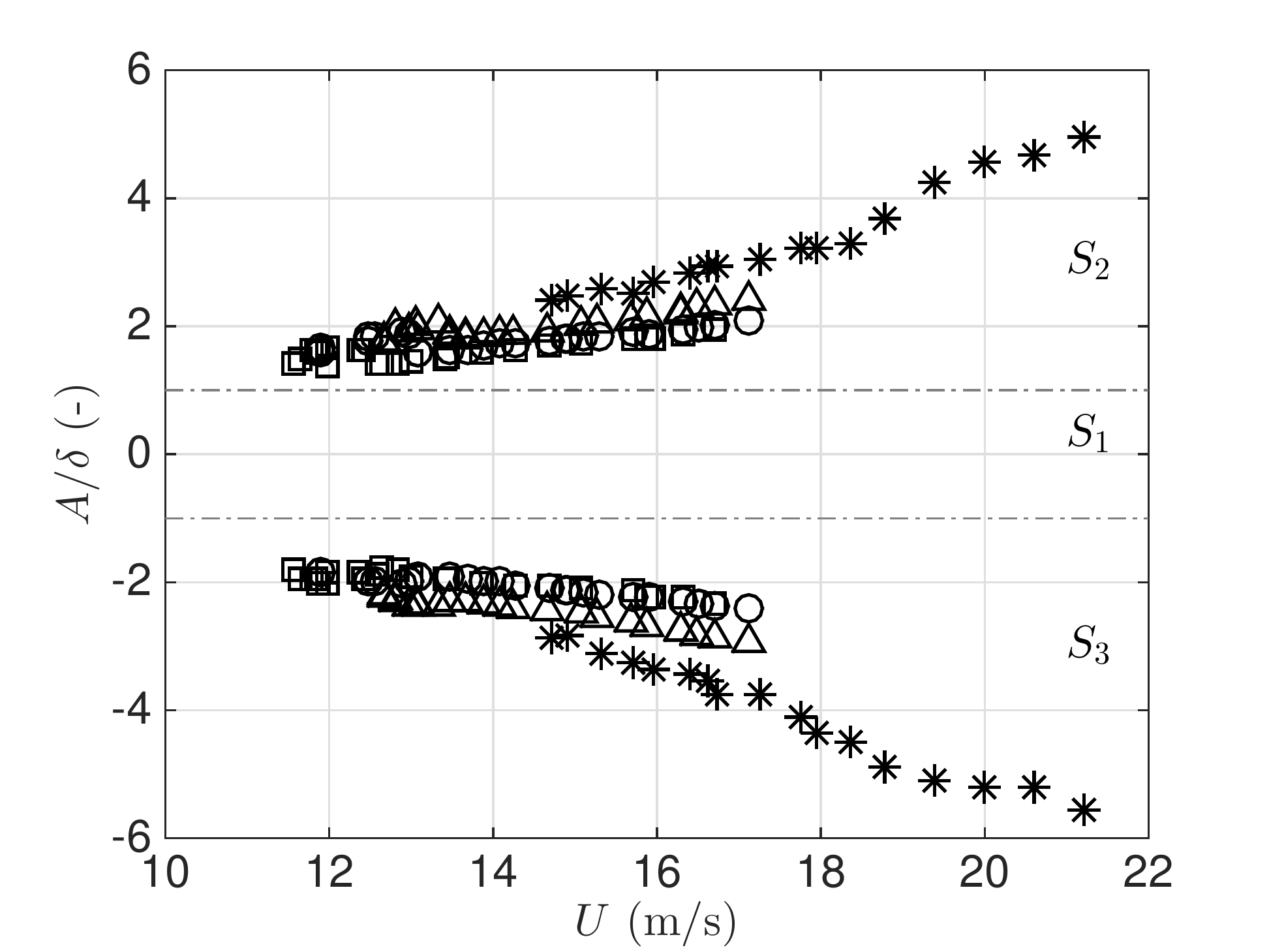}}
  \subfigure[Frequency]{\label{bifF_AOA0} \includegraphics[height=.35\textwidth]{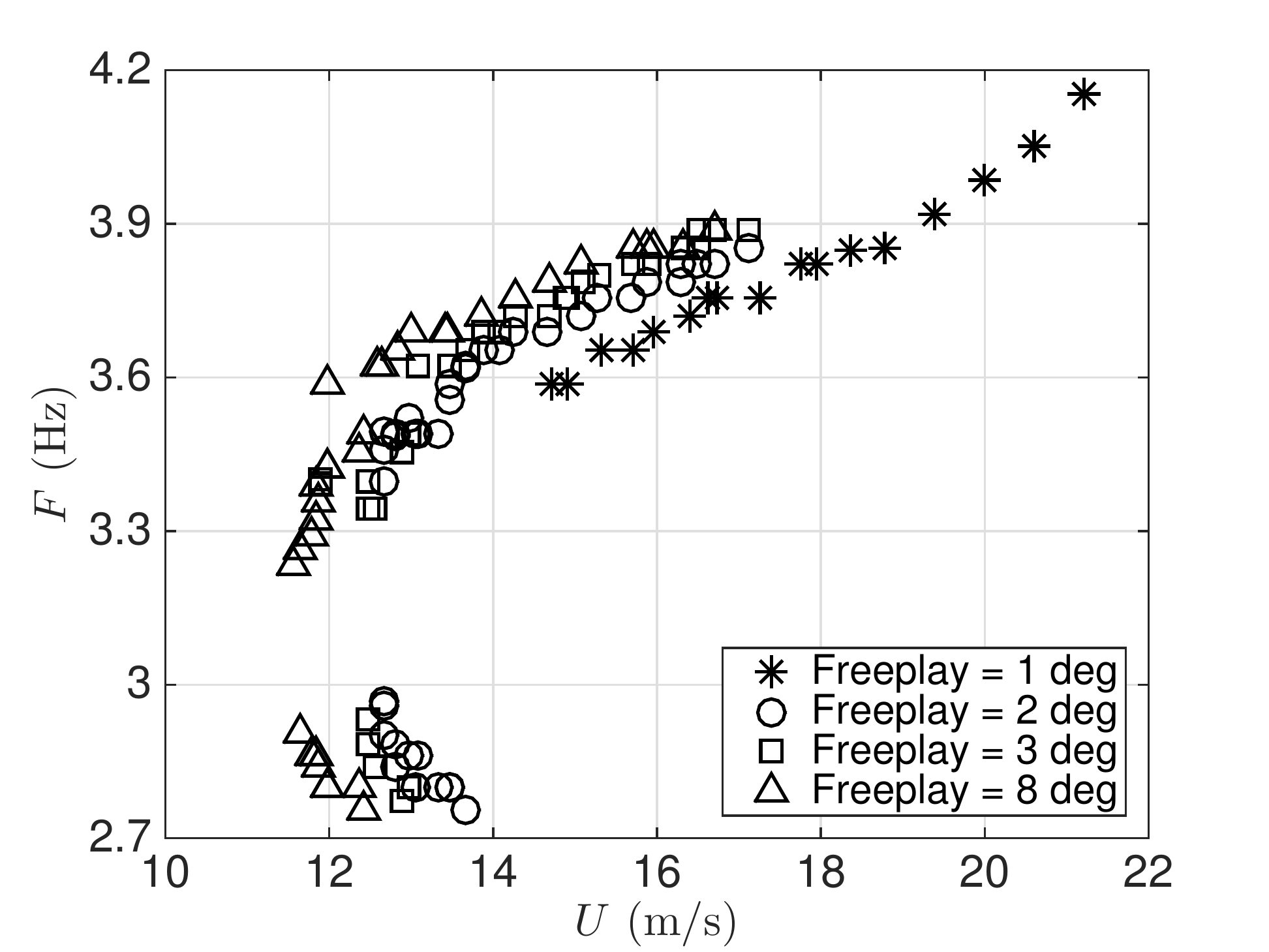}}
  \end{center}
  \caption{Bifurcation diagram of the system with $\alpha_p = 0$}
\label{bif_AOA0}
\end{figure}

\begin{figure}[ht]
  \begin{center}
  \subfigure[U = 11.8 m/s]{\label{TS5} \includegraphics[height=.35\textwidth]{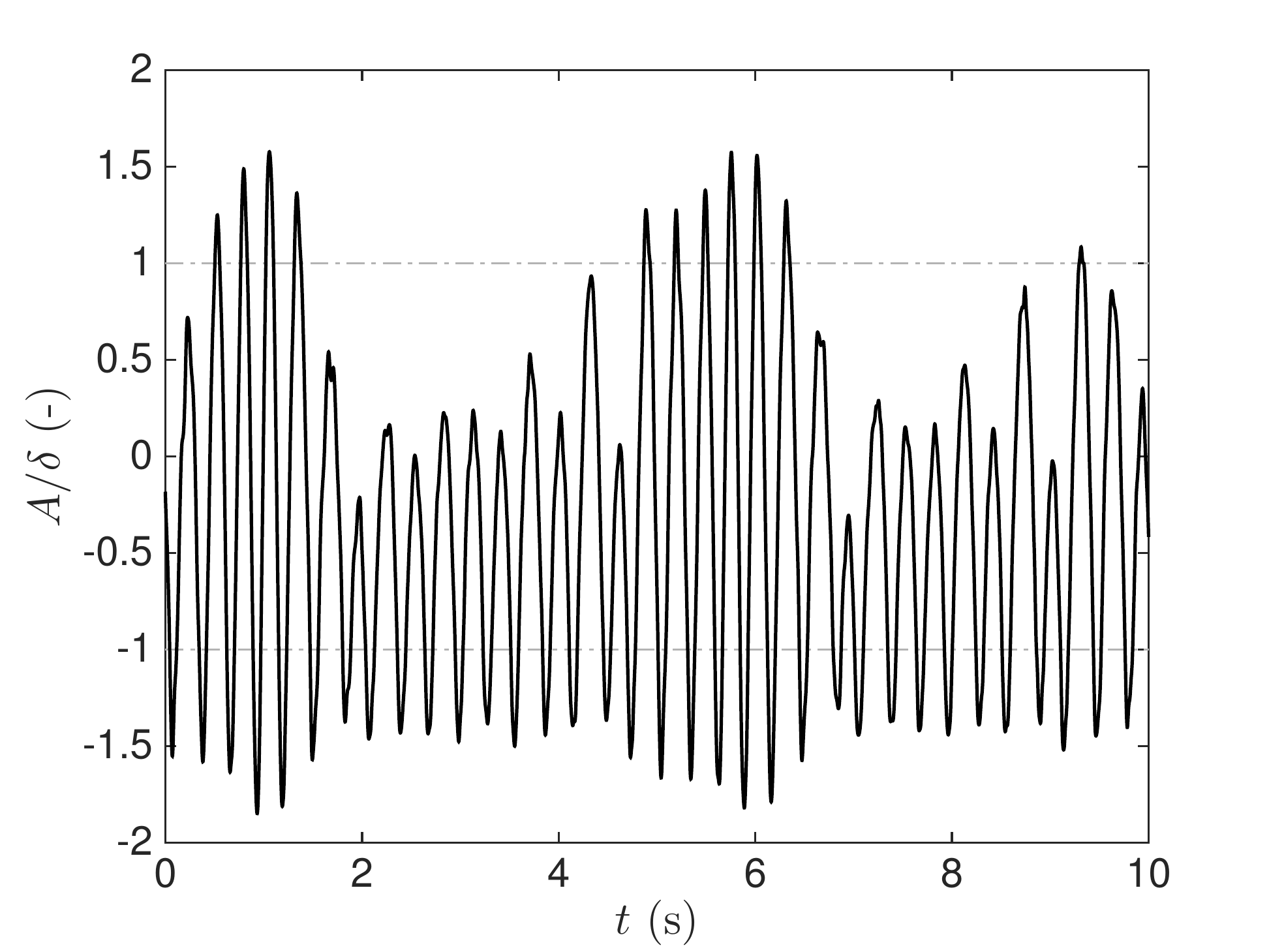}}
  \subfigure[U = 12.5 m/s]{\label{TS6} \includegraphics[height=.35\textwidth]{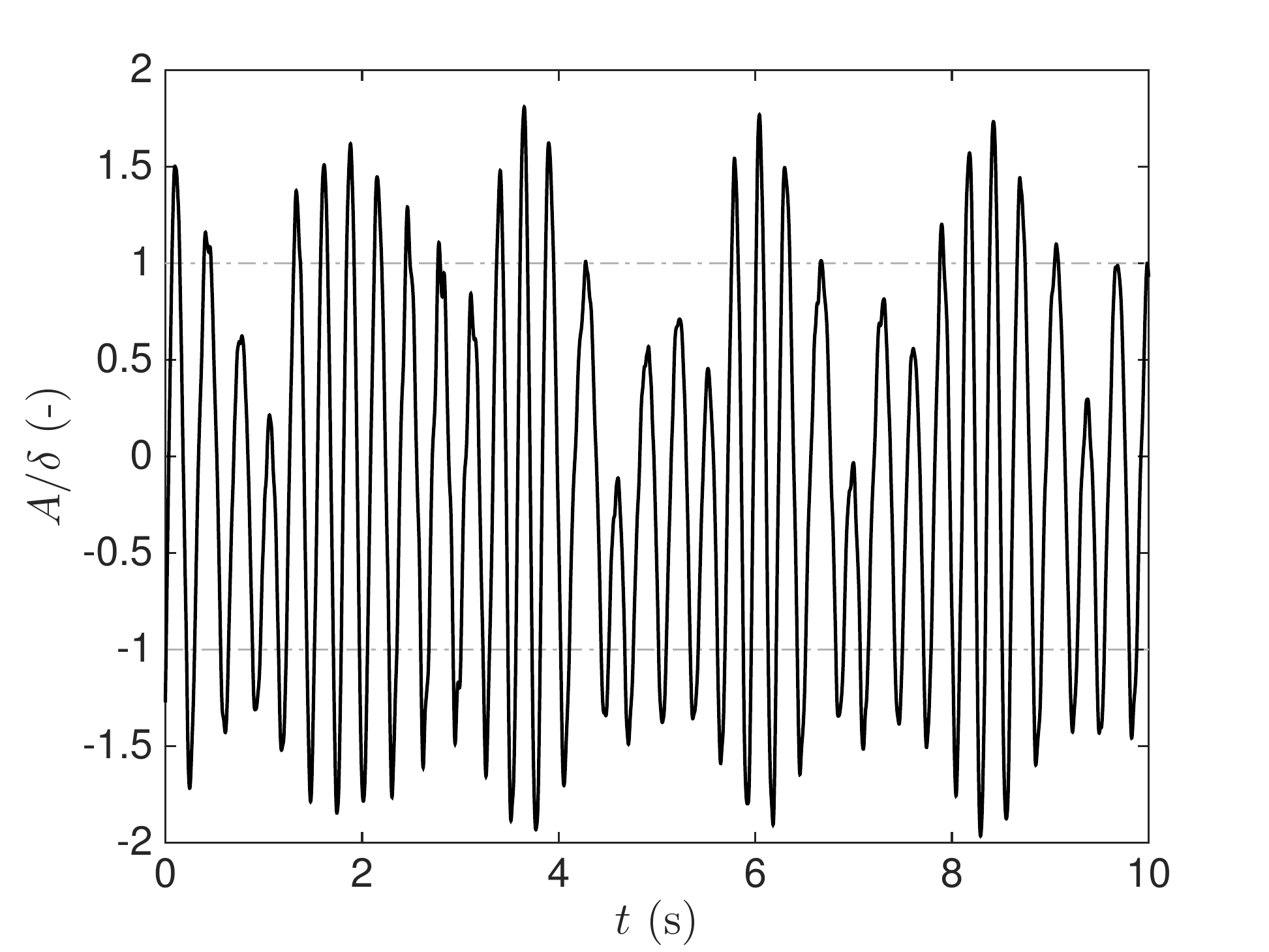}}
  \subfigure[U = 12.9 m/s]{\label{TS7} \includegraphics[height=.35\textwidth]{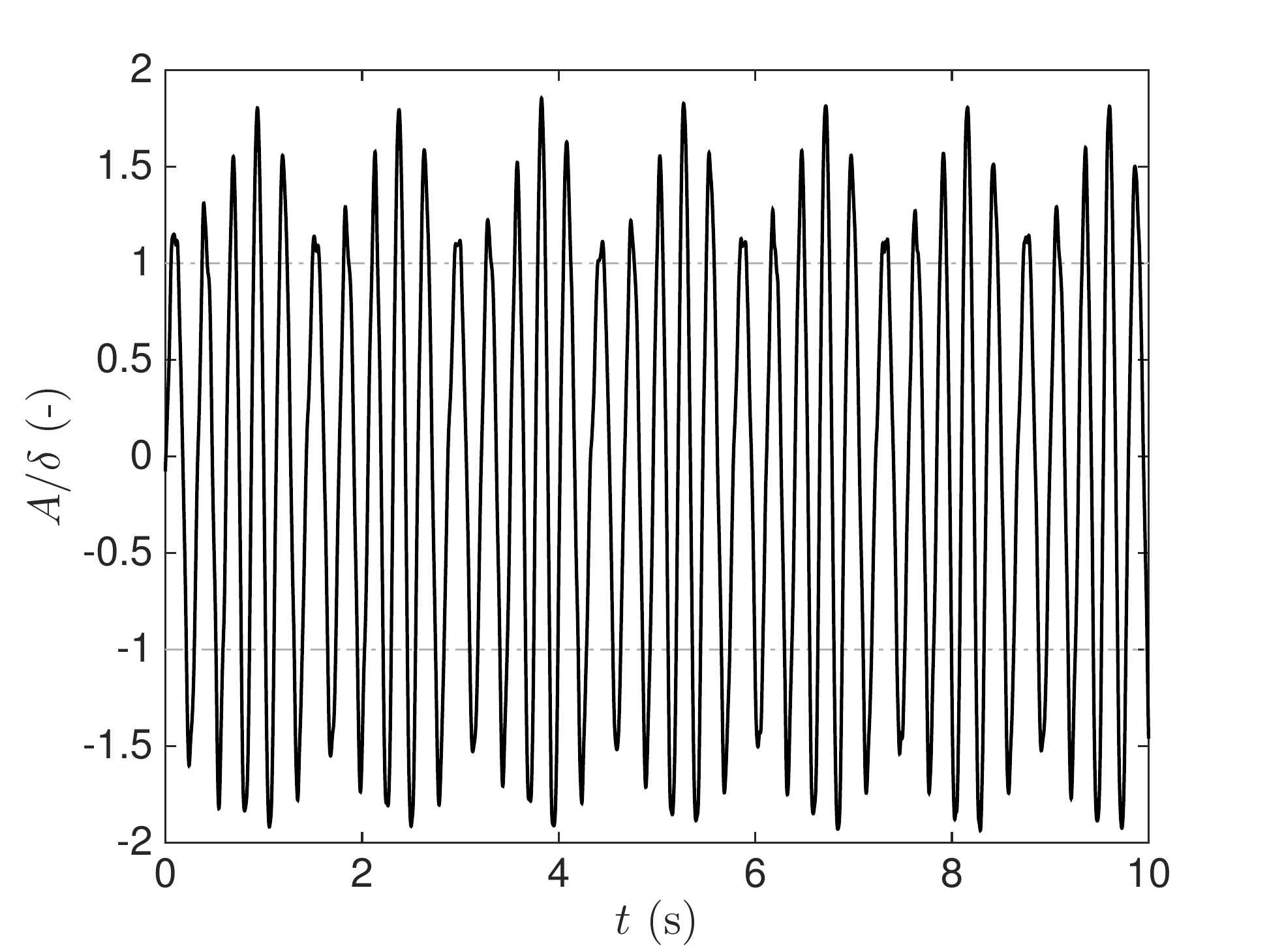}}
  \subfigure[U = 13.1 m/s]{\label{TS8} \includegraphics[height=.35\textwidth]{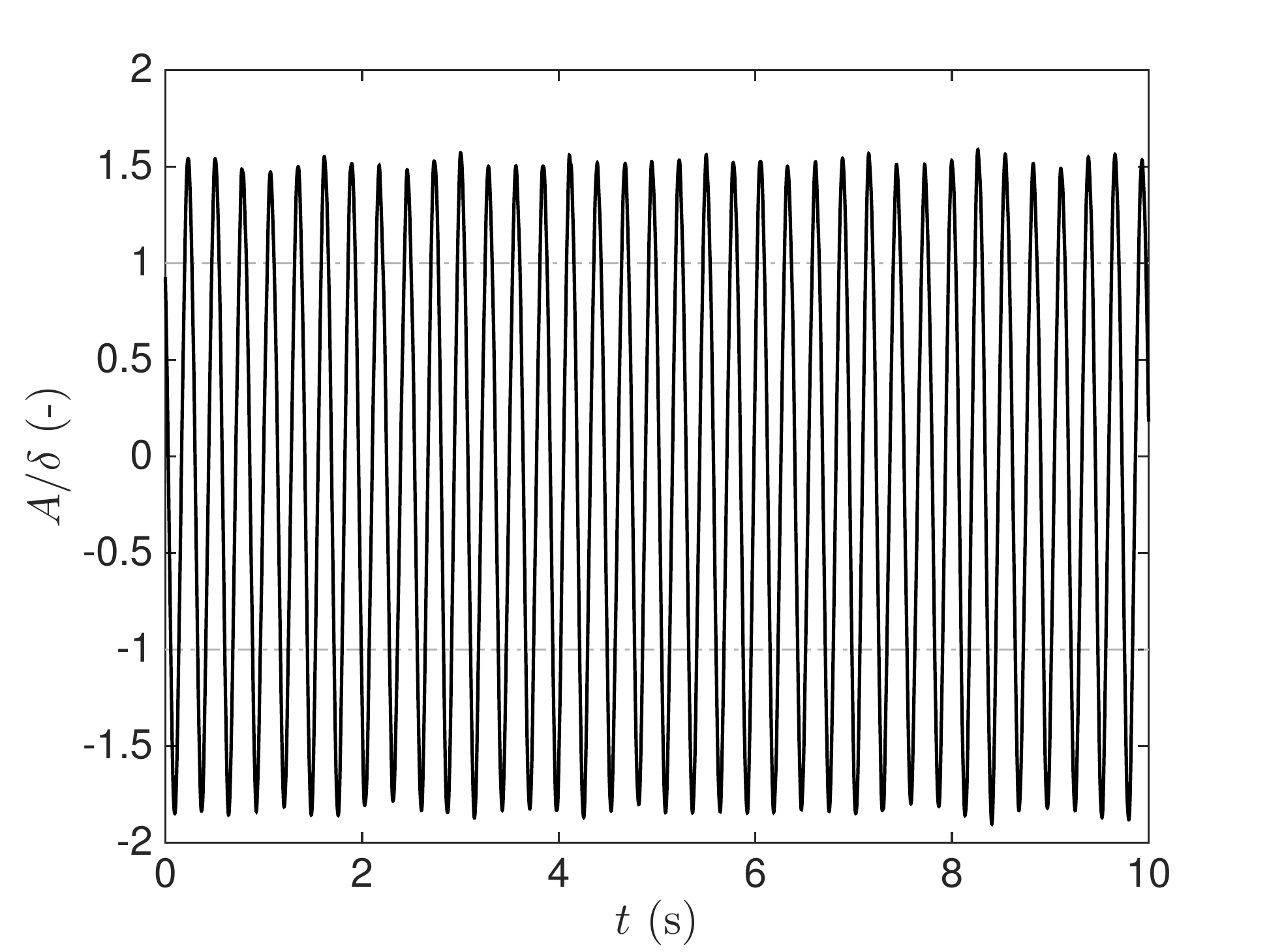}}
  \end{center}
\caption{Pitch time history response of the system with freeplay = 3 deg and $\alpha_p = 0$ deg at airspeeds of 11.8, 12.5, 12.9 and 13.1 m/s}
\label{TS_exp1}
\end{figure}

\begin{figure}[ht]
  \begin{center}
  \subfigure[Amplitude]{\label{bifA_AOA5} \includegraphics[height=.35\textwidth]{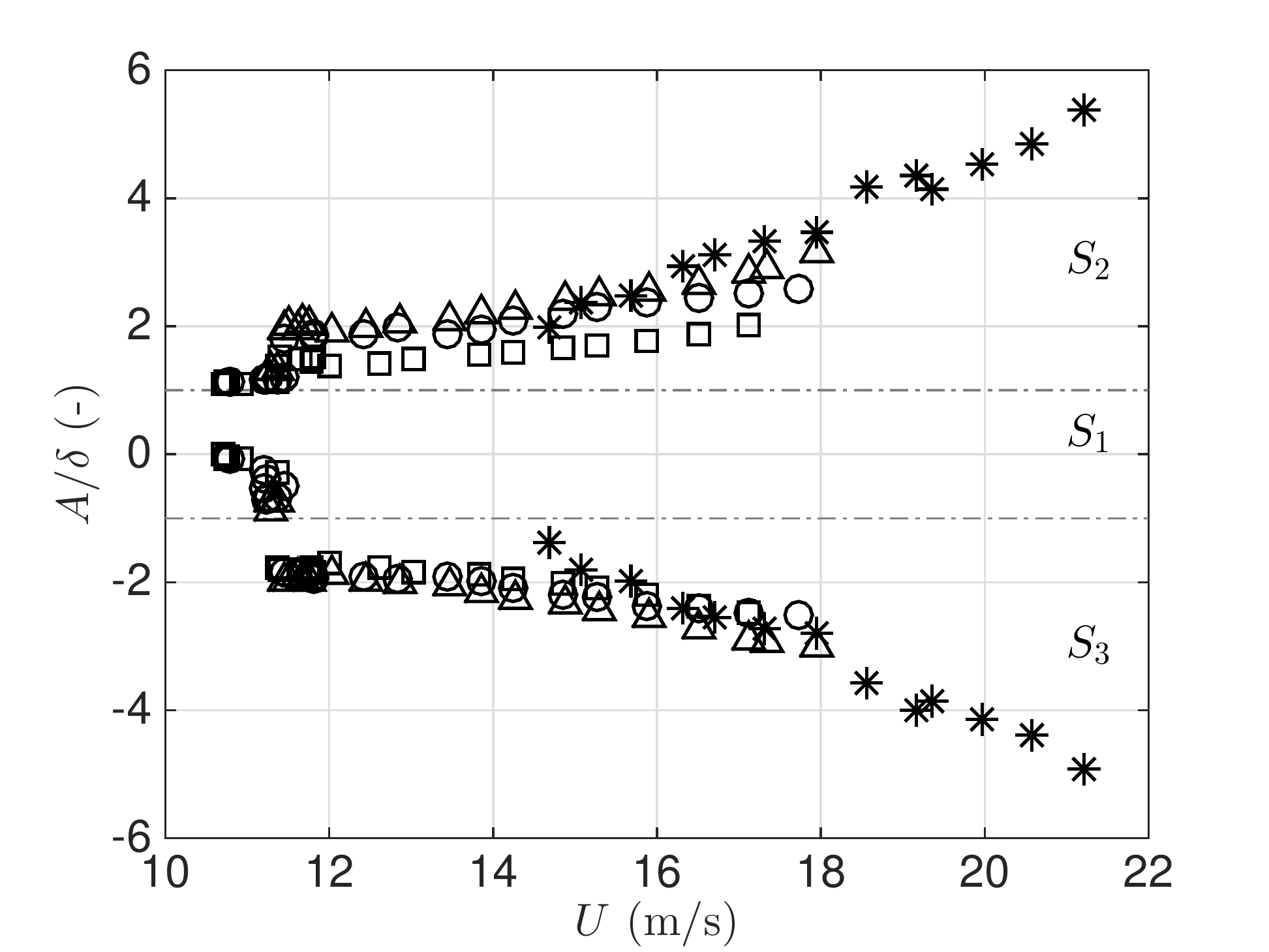}}
  \subfigure[Frequency]{\label{bifF_AOA5} \includegraphics[height=.35\textwidth]{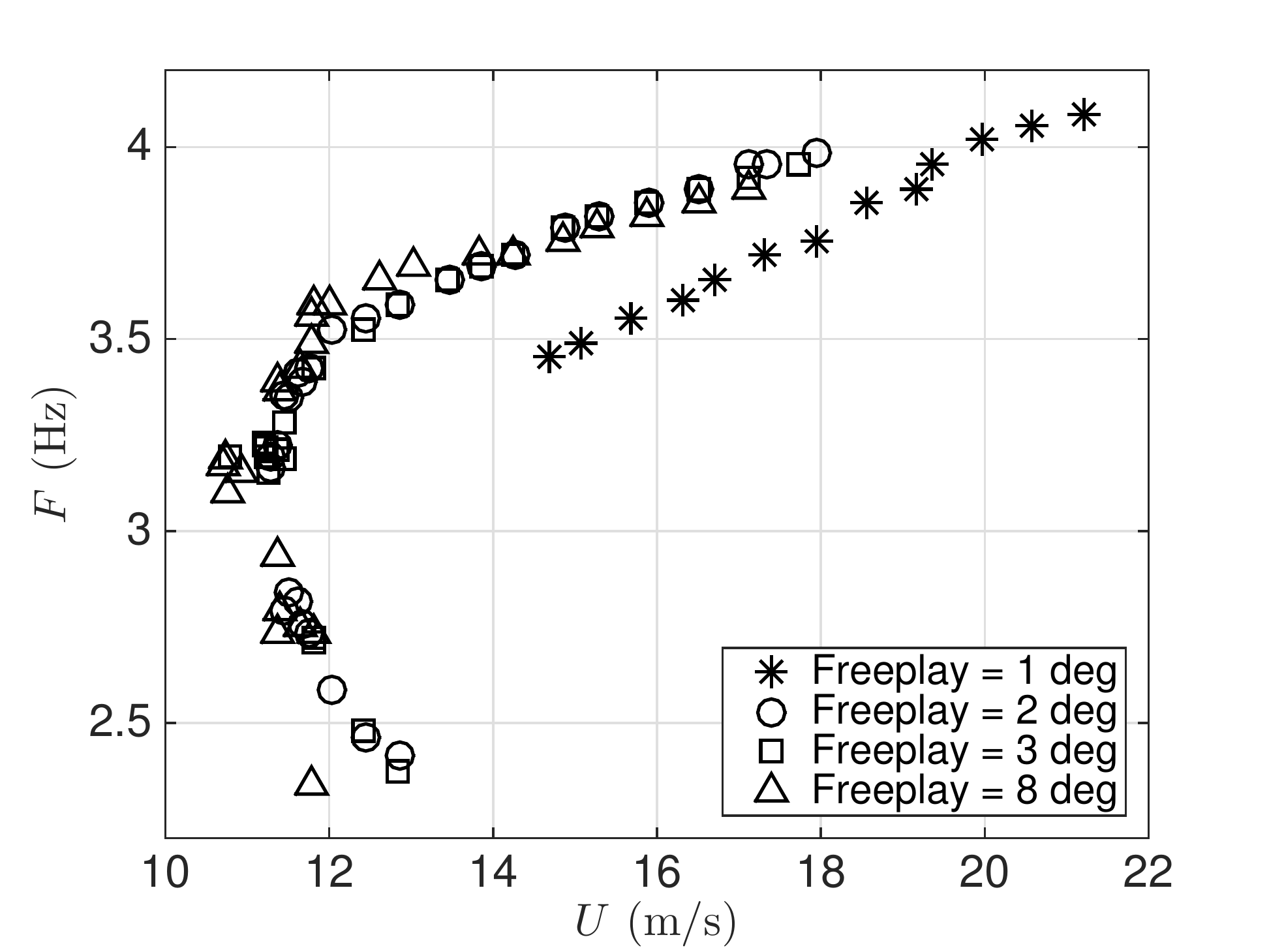}}
  \end{center}
  \caption{Bifurcation diagram of the system with $\alpha_p = 5$ deg}
\label{bif_AOA5}
\end{figure}

\begin{figure}[ht]
  \begin{center}
  \subfigure[U = 10.7 m/s]{\label{TS1} \includegraphics[height=.35\textwidth]{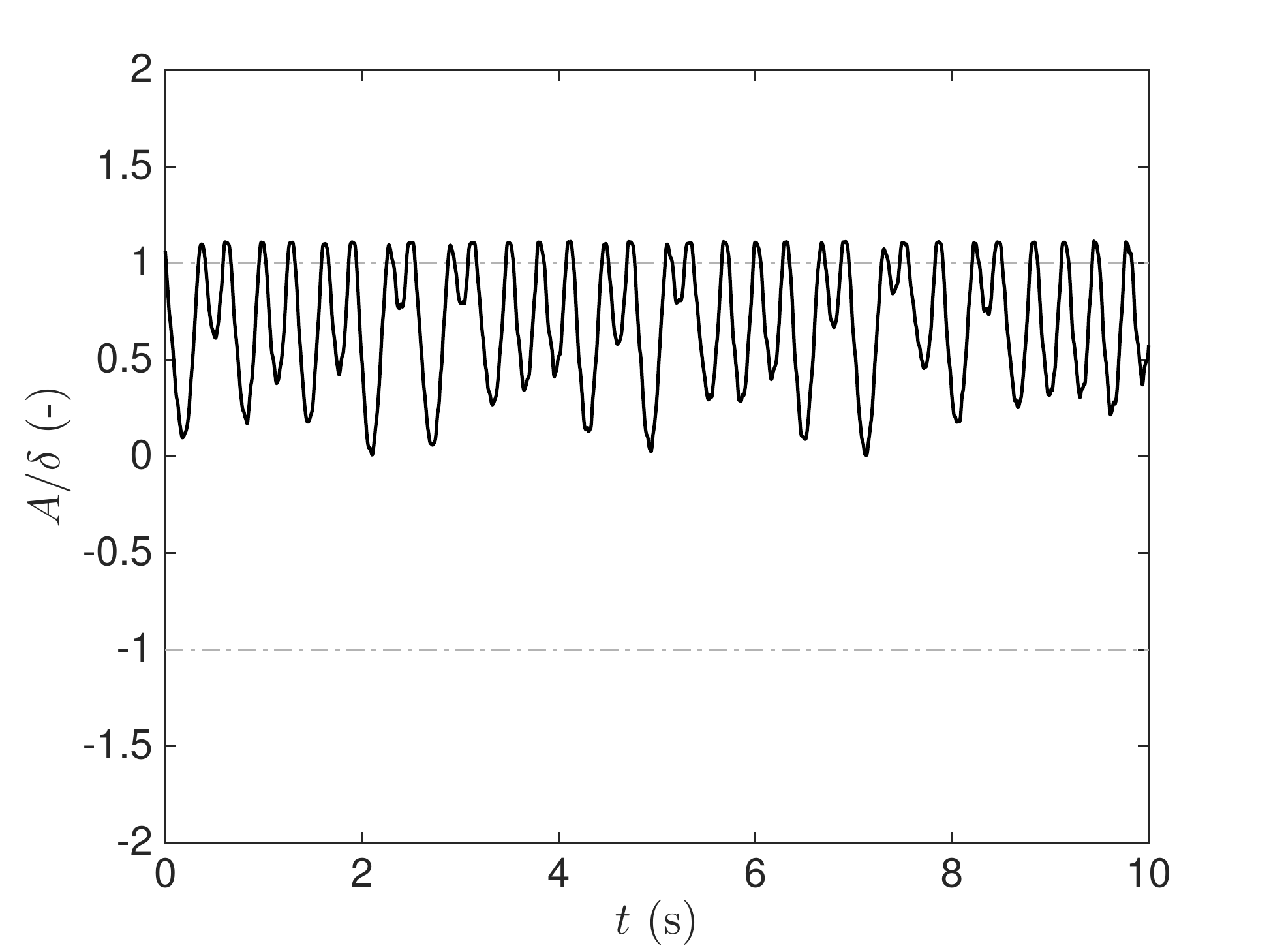}}
  \subfigure[U = 11.3 m/s]{\label{TS2} \includegraphics[height=.35\textwidth]{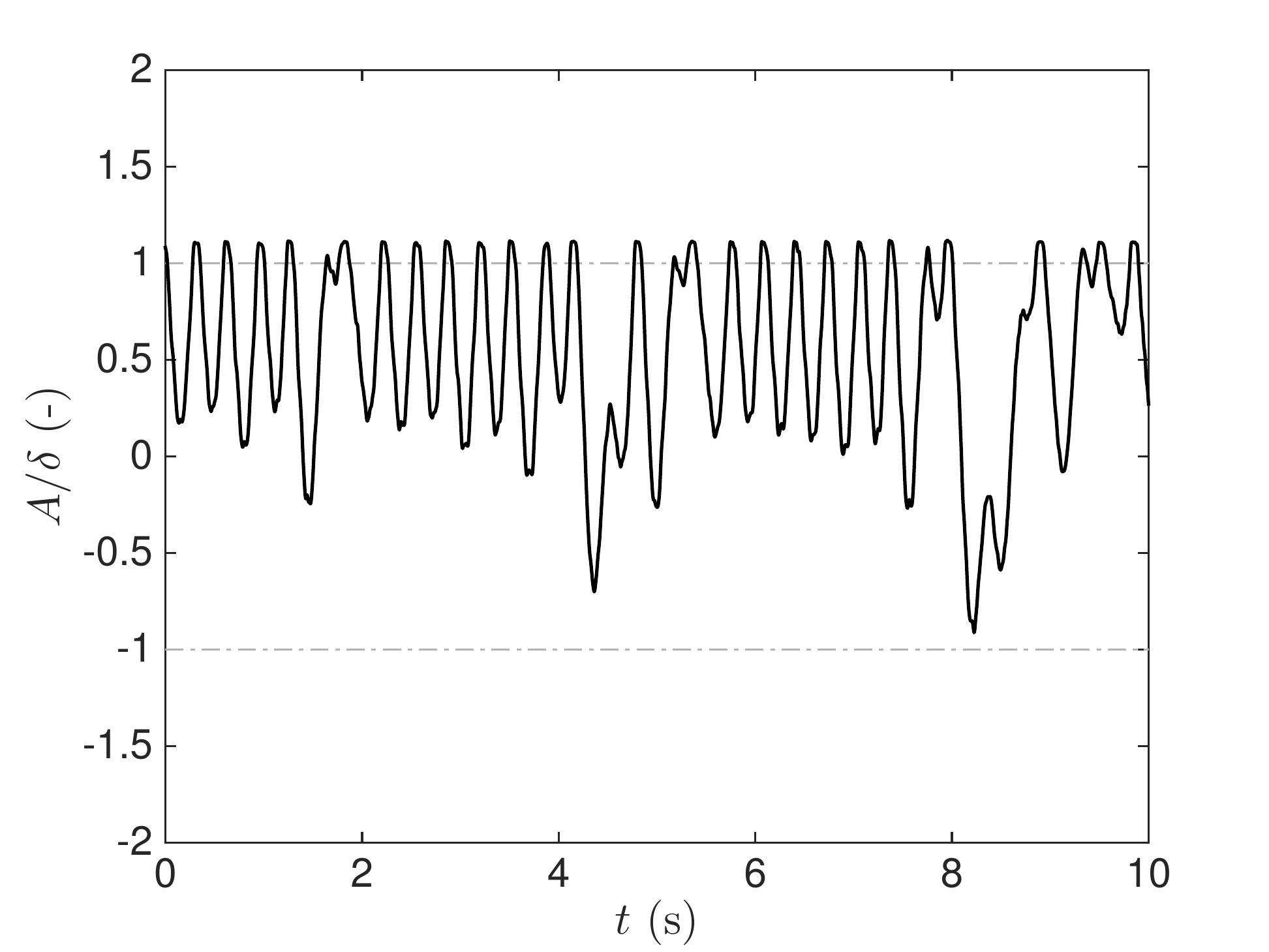}}
  \subfigure[U = 11.3 m/s]{\label{TS3} \includegraphics[height=.35\textwidth]{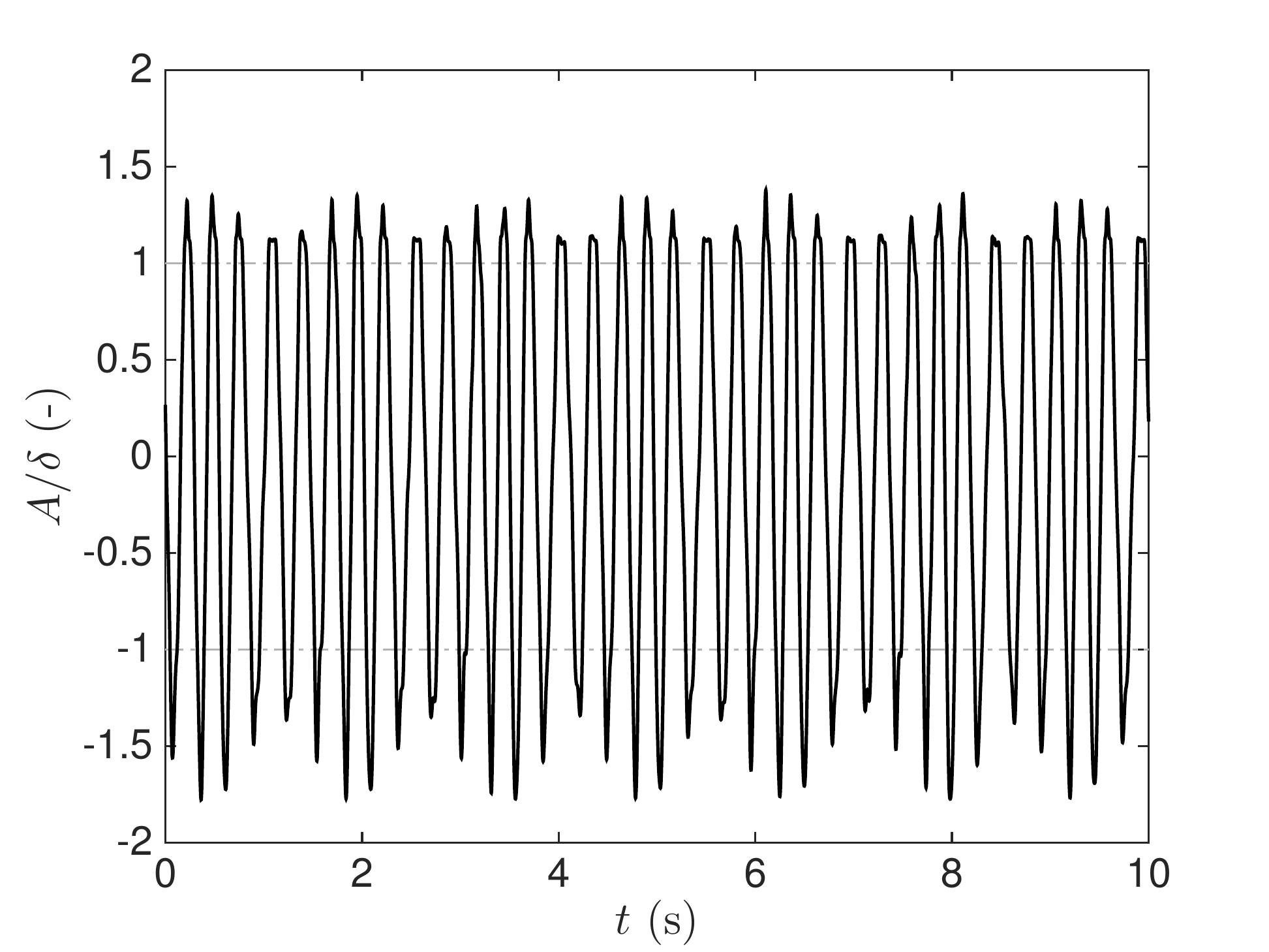}}
  \subfigure[U = 11.8 m/s]{\label{TS4} \includegraphics[height=.35\textwidth]{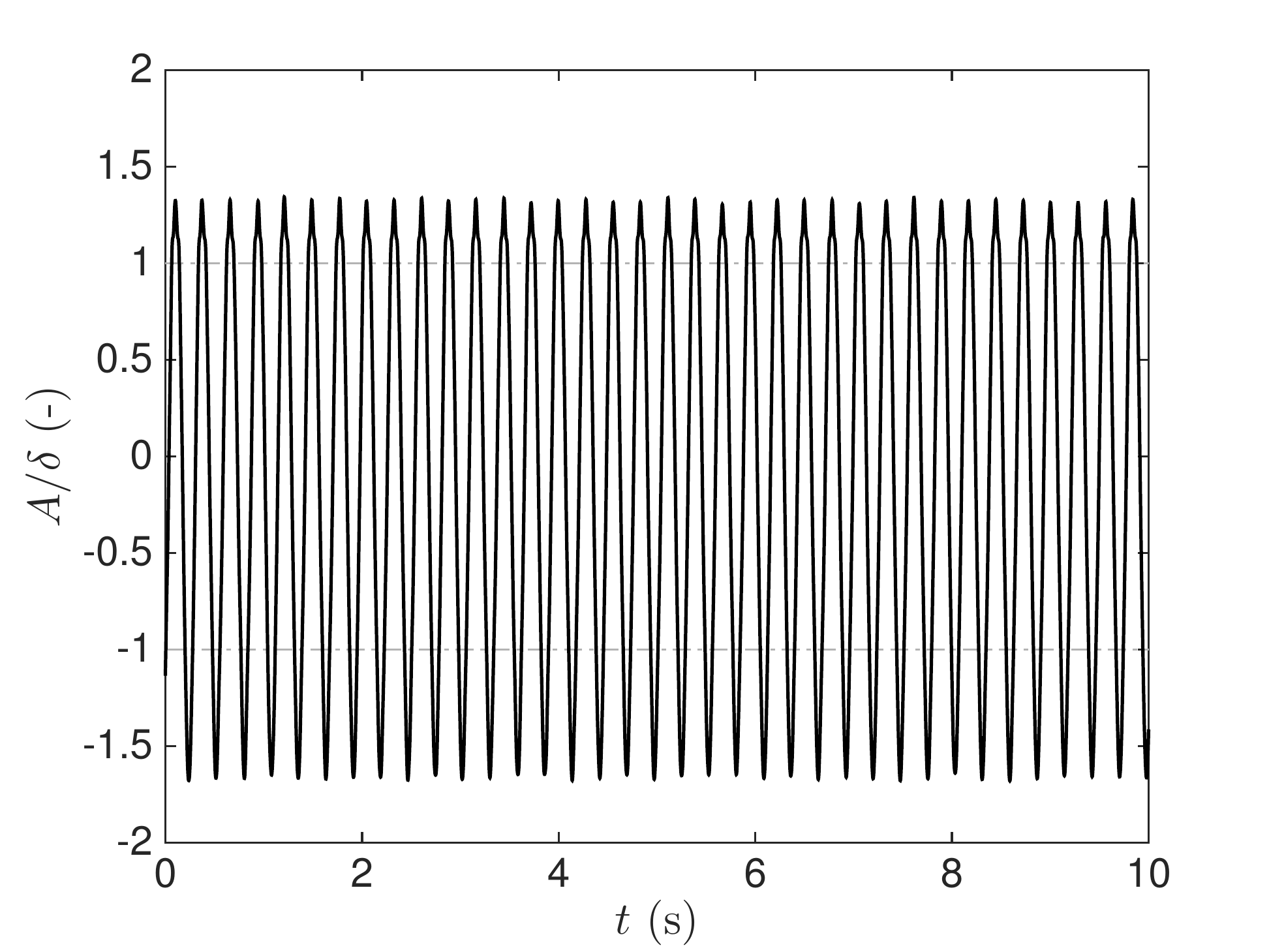}}
  \end{center}
\caption{Pitch time history response of the system with  freeplay = 8 deg and $\alpha_p = 5$ deg at airspeeds of 10.7, 11.3, 11.3 and 11.8~m/s}
\label{TS_exp2}
\end{figure}

\revA{Figures \ref{bifA_AOA0} and \ref{bifF_AOA0} plot the pitch amplitude and frequency of the limit cycles obtained without aerodynamic preload ($\alpha_p \approx 0$) for all airspeeds and freeplay cases}. The largest three freeplay values (2, 3 and 8 deg) exhibit similar behaviour as all the amplitude and frequency results fall on a single curve for each of the three freeplay values, as already highlighted by numerous previous studies on freeplay. For the smallest freeplay \revA{gap} (stars), the results \revA{are slightly different}. In this case, the amplitude ratio and the LCO onset speed are \revA{higher} than in the other cases and the LCO frequency is slightly \revA{lower}. These differences are attributed to the nonlinear friction in the bearings and to the geometry of the freeplay gap. \revA{Marsden and Price observed a similar effect on a pitch-plunge system with freeplay and bearings in the pitch~\cite{marsden1}.} In all four cases, the system undergoes a slow and almost linear amplitude increase with airspeed once the oscillations have started. The LCO frequency features a main branch that starts at about 3.3~Hz and increases to up to 4.2~Hz. The points with frequencies under 3 Hz arise from a secondary peak observed in the Fast Fourier \revA{Transforms} (FFT) of quasi-periodic oscillations. 
\LLL 
The amplitude bifurcation diagram presented in fig.~\ref{bifA_AOA0} plots the peak-to-peak amplitude of the measured response signals and misses an important part of the dynamics: the existence of two-domain and three-domain LCOs. Figure \ref{TS_exp1} plots pitch time history responses of the system with a \revA{freeplay gap} of 3~deg and without aerodynamic preload. At 11.8~m/s, the system undergoes quasi-periodic limit cycle oscillations. The responses switch from two-domain to three-domain oscillations and vice-versa in a \revA{quasi-periodic} fashion because the system is attracted by both solutions. Increasing the airspeed to 12.5~m/s leads to a motion dominated by three-domain oscillations with infrequent occurrences of two-domain oscillations. Then at 12.9~m/s, the two-domain oscillations completely disappear and only three-domain quasi-periodic oscillations are observed up to 13.1~m/s where the system undergoes mono-harmonic limit cycles. \revA{Figure \ref{TS_exp1} clearly shows} that two-domain cycles become less and less frequent as the airspeed is increased until they totally disappear and only three-domain mono-harmonic LCOs are observed.  
\LLL
The bifurcation diagrams of the system with an aerodynamic preload angle $\alpha_p$ of 5 deg are shown in figure \ref{bif_AOA5}. Once again, the largest 3 freeplay values lead to similar results while the measurements from the case with the smallest freeplay \revA{are slightly different}. This time, two-domain LCOs are clearly observed. They appear at around 11m/s and occur continuously, not intermittently as was the case for $\alpha_p=0$ deg.  \revA{Their} amplitude smoothly increases until \revA{approximately $U = 12$~m/s}, where they are replaced by three-domain LCOs. The frequency diagram is similar to the one obtained without preload, the main LCO branch undergoes a frequency increase with airspeed while another branch\revA{, corresponding to the secondary peak of the FFT of quasi-periodic oscillations}, features a decreasing frequency variation with airspeed. 
\LLL
Figure \ref{TS_exp2} highlights the transition from two-domain LCOs to three-domain LCOs with a freeplay of 8~deg and a preload angle of 5~deg. At 10.7~m/s, a clear two-domain LCO is observed. Its \revA{positive peak amplitude} is constant \revA{in time} while its negative peak \revA{amplitude varies unpredictably from cycle to cycle}. At 11.3~m/s, two responses can be observed, depending on the initial condition. The first one, plotted in figure \ref{TS2}, is another two-domain limit cycle with a larger negative amplitude than in the lower-speed case. The second one, depicted in figure \ref{TS3}, is a quasi-periodic LCO where all the periods span the three domains. Finally, at 11.8~m/s only mono-harmonic three-domain LCOs are observed. 
\LLL 
Even \revA{though} no clear two-domain LCOs were observed without aerodynamic preload, the system exhibited a smooth transition from a motion dominated (or influenced) by the two-domain LCOs to a purely three-domain LCOs motion where a single dominant harmonic was observed. 
\clearpage
\section{Mathematical model of the experiment} 

\begin{figure}[ht]
  \begin{center}
  \subfigure[Pitch damping ratio]{\label{flutter_damp_pi} \includegraphics[height=.35\textwidth]{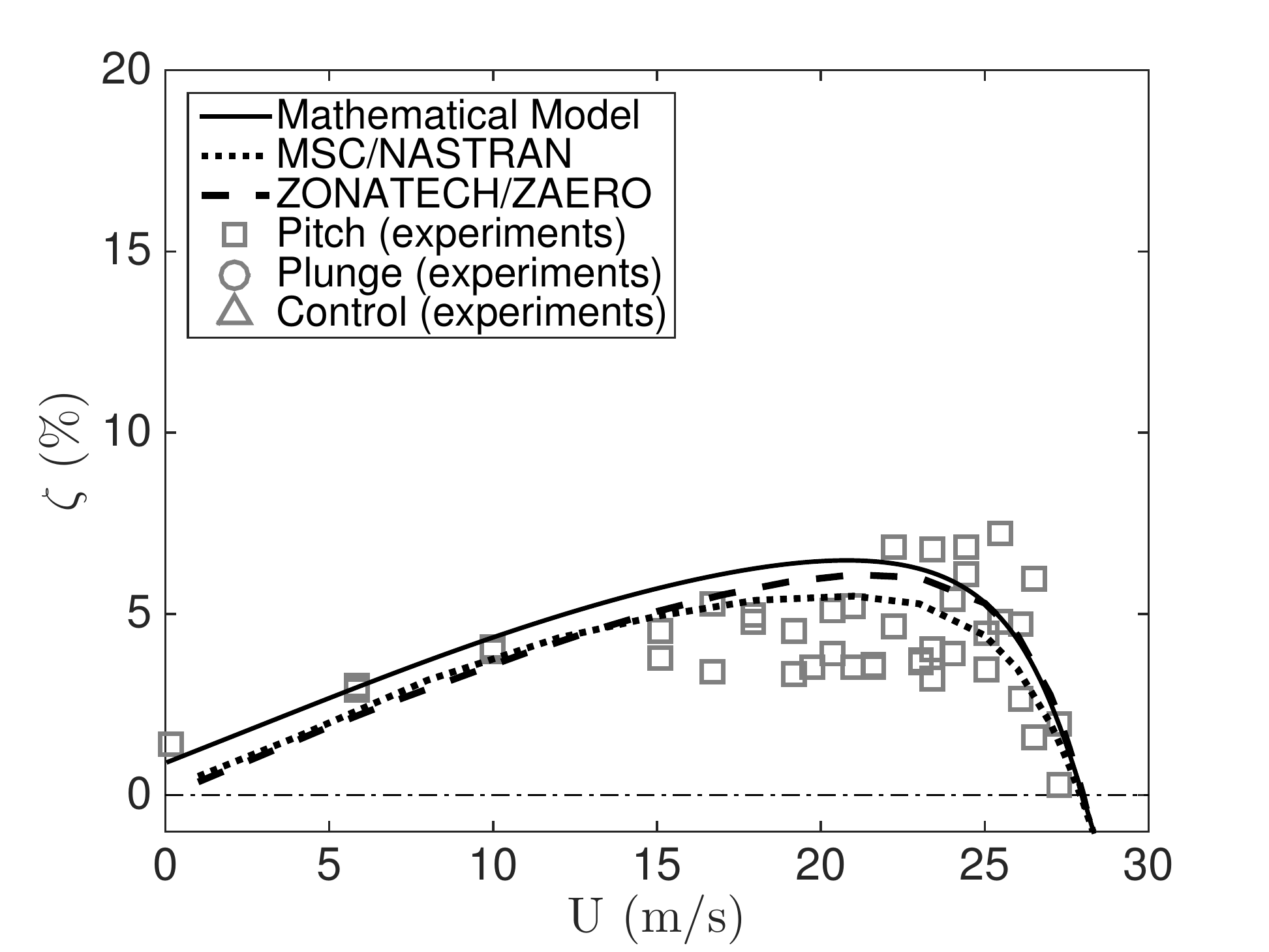}}
  \subfigure[Plunge damping ratio]{\label{flutter_damp_pl} \includegraphics[height=.35\textwidth]{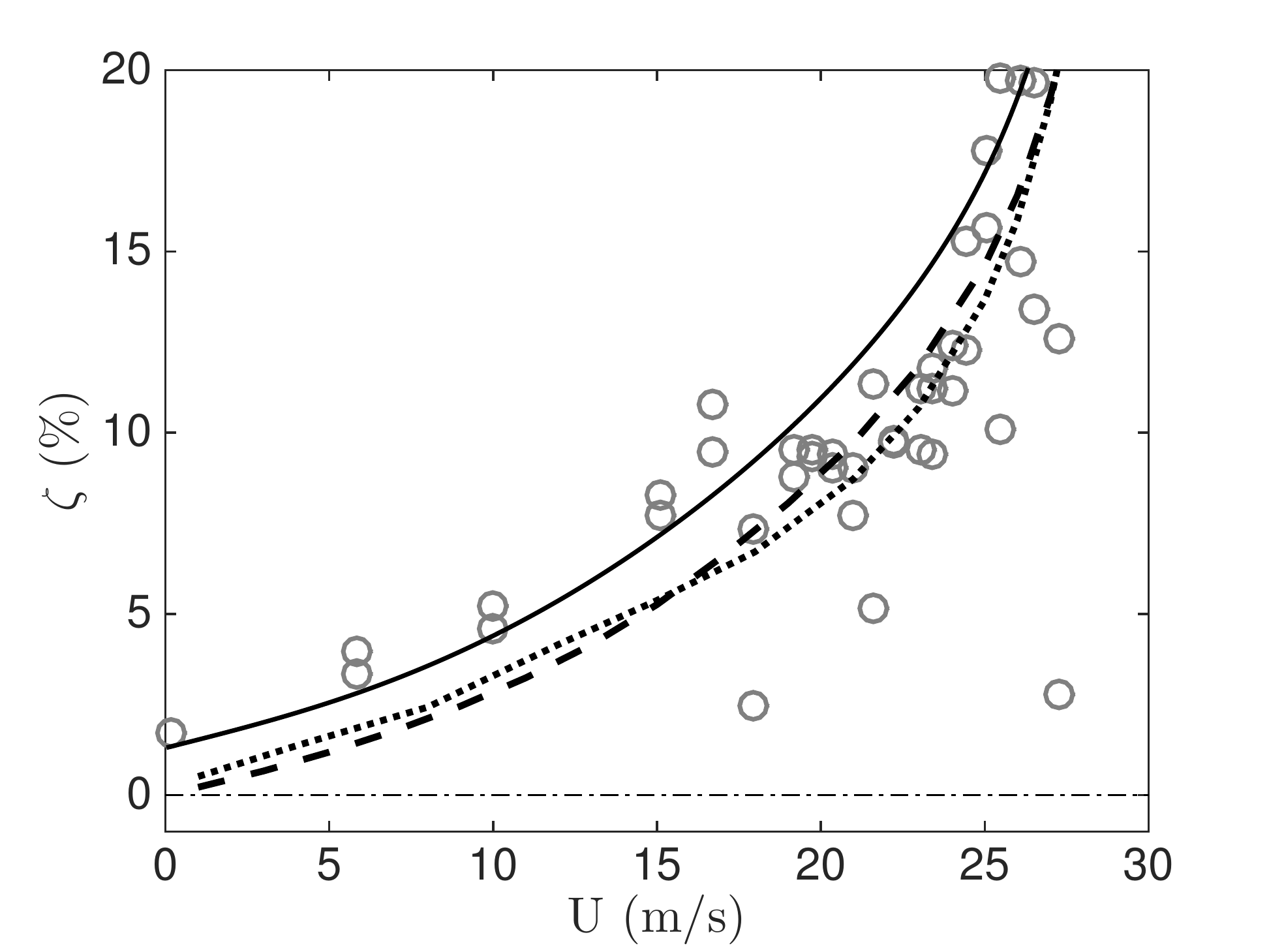}}
  \subfigure[Control damping ratio]{\label{flutter_damp_co} \includegraphics[height=.35\textwidth]{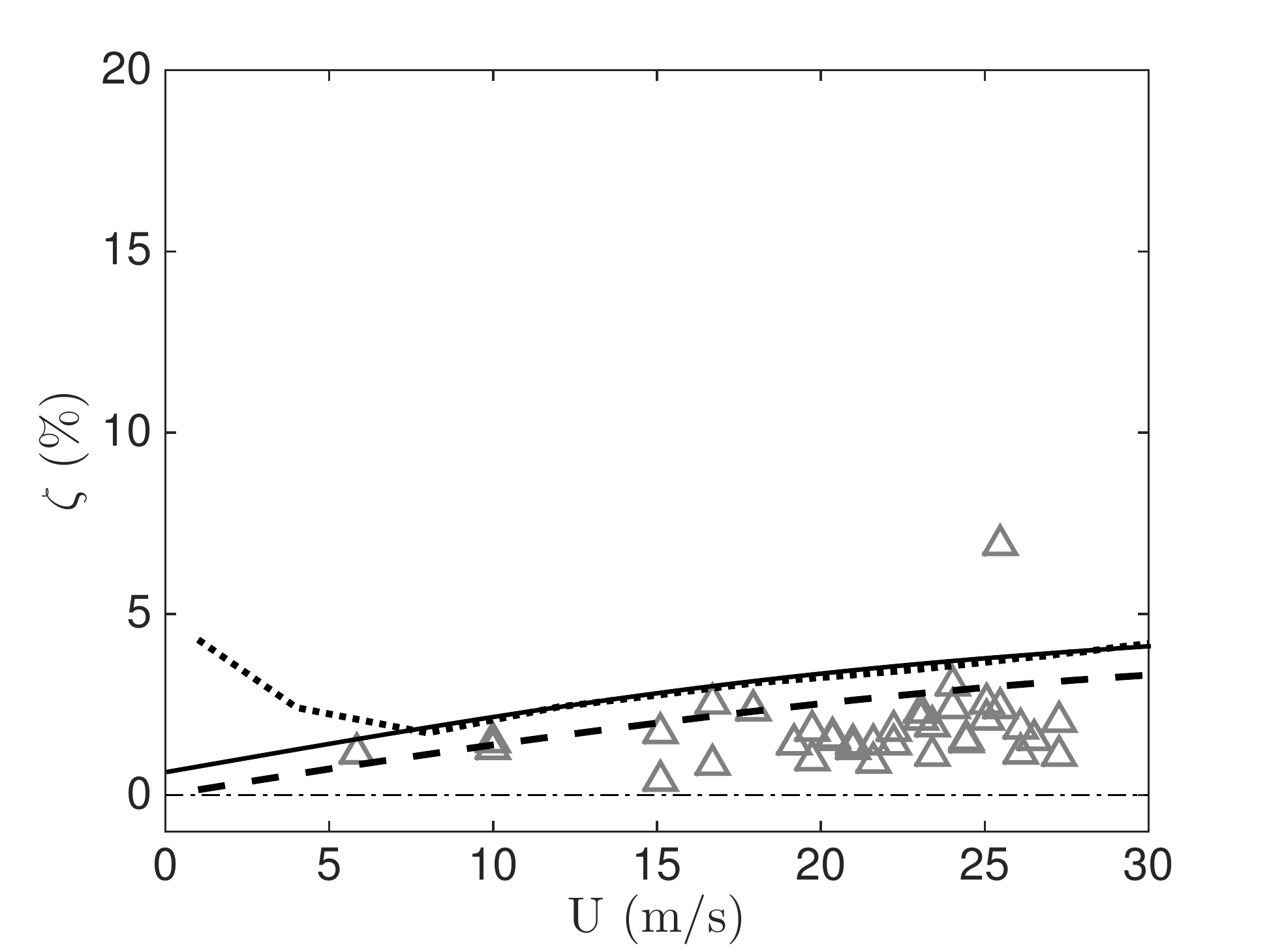}}
  \subfigure[System natural frequencies]{\label{flutter_freq} \includegraphics[height=.35\textwidth]{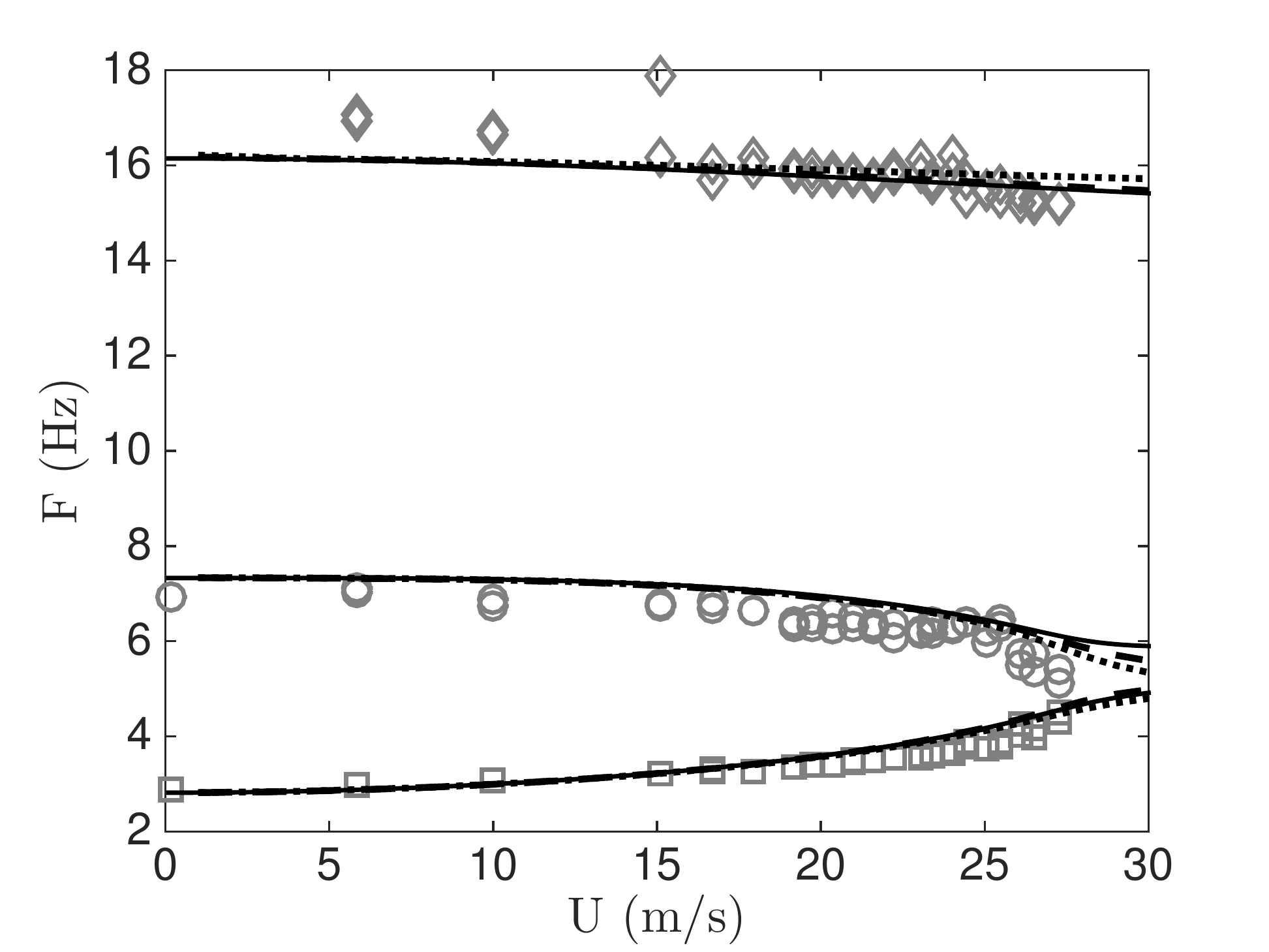}}
  \end{center}
  \caption{\revA{Flutter plots for overlying linear system. The legend in subfigure 1 applies to all 4 subfigures.}}
\label{flutter_freq_damp}
\end{figure}

The primary mathematical model of the experiment was based on equations~\ref{eq_dof3_23}. While care was taken to make the experiment as symmetric as possible (aside from the aerodynamic preload), neither the wing nor its placement in the tunnel were perfectly symmetric. In particular, it was impossible to ensure that the wing was perfectly vertical, which means that gravity played a small role in introducing additional asymmetry. The mathematical model of the experiment represented all the sources of asymmetry as a gravity effect due to imperfect verticality. Therefore, a gravity term was added to equation~\ref{eq_dof3_23}\revA{; for} a constant roll angle $\theta$, the pitch torque due to gravity is written as
\begin{equation}
T_{grav} = \sin(\theta) \cos(\alpha) S g \approx \sin(\theta) (1 - \frac{\theta^2}{2}) S \approx \sin(\theta) S g
\end{equation}
where $S = (x_{cg} - x_f) m$ is the static imbalance of the wing and $g$ is the acceleration due to gravity. Assuming small pitch angles and \revA{neglecting second} order terms yields a constant torque that depends on the roll angle of the setup. The constant moment of force is then multiplied by $\mathbf{q}_n$ and added to equation \ref{eq_dof3_23} to obtain the full equations of motion of the model. 
\revA{
\begin{equation}
\dot{\mathbf{x}} = \left\{
\begin{array}{ll}
\mathbf{Q}_1\mathbf{x}+\mathbf{q}_p \alpha_p + \mathbf{q}_n T_{grav} & \mbox{\hspace{2ex} if \hspace{2ex}} |\alpha| \leq \delta  \\ \mathbf{Q}_2\mathbf{x}-\mathbf{q}_n K_{\alpha} \sgn(\alpha) \delta+\mathbf{q}_p \alpha_p + \mathbf{q}_n T_{grav}  & \mbox{\hspace{2ex} if \hspace{2ex}} |\alpha| > \delta
\end{array}
\right.
\label{eq_dof3_23_wgrav}
\end{equation}}
The values of the structural parameters of the equations of motion are those given in appendix B. An equivalent mathematical model representing the reference analytical solution was created in MSC/NASTRAN~\cite{nastran} for comparison. Pure plunge, pitch and control modes were enforced following the reasoning of the ZONATECH/ZAERO model given in \cite{ZAERO_AM2}. The aerodynamic model consists of \revA{doublet lattice} panels distributed along a very long span wing. Figure~\ref{flutter_freq_damp} plots the variation of the natural frequencies and damping ratios of the overlying linear system with airspeed. \revA{Modal estimates from the wind tunnel tests} are compared to the predictions of the three models. Clearly the predictions of the models are in good agreement with each other and with the experimental measurements. 

\section{Bifurcation analysis using equivalent linearisation} 

\begin{figure}[ht]
  \begin{center}
    \includegraphics[height=.45\textwidth]{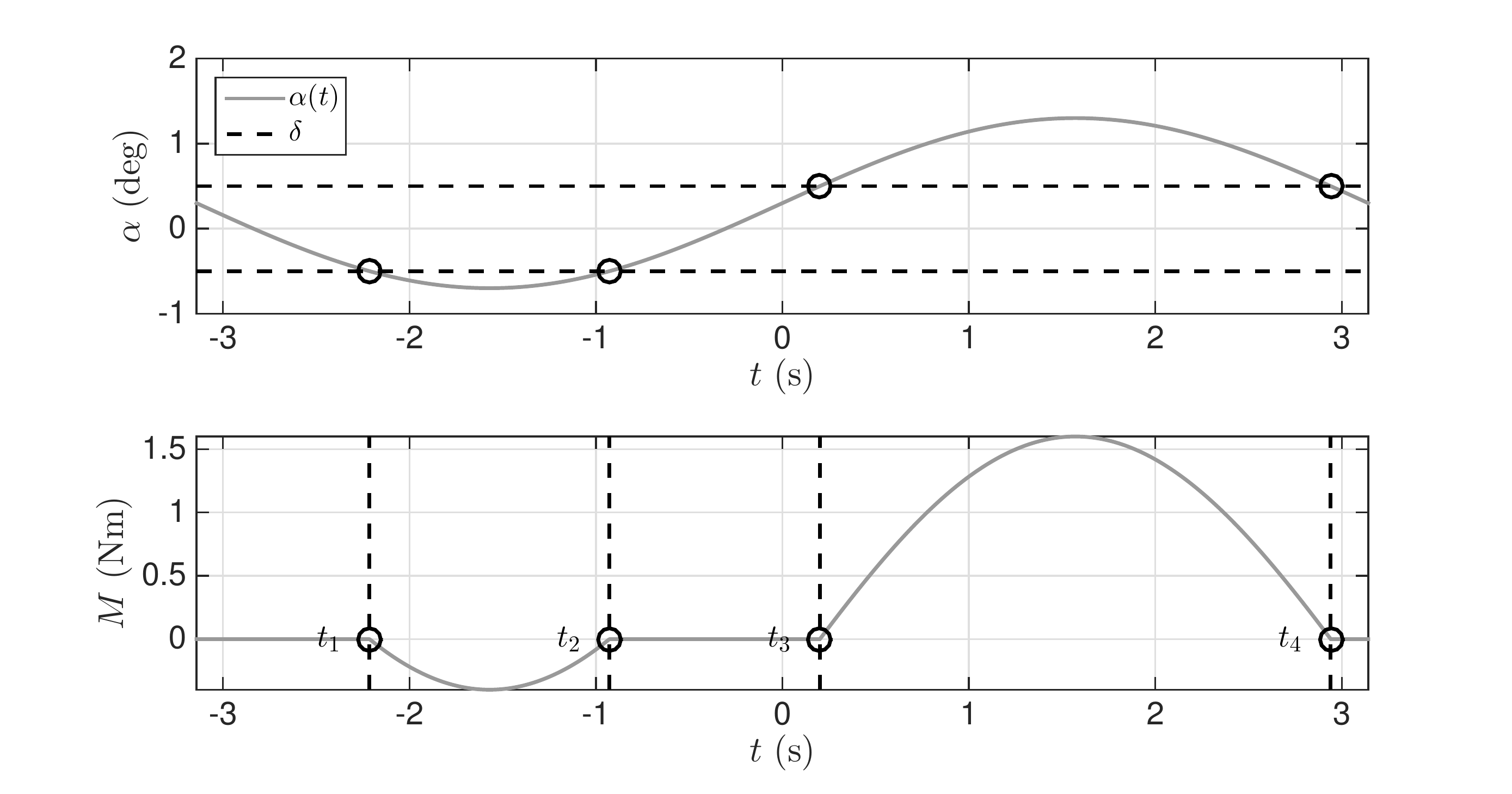}
  \end{center}
  \caption{Sinusoidal displacement (top) and corresponding freeplay load (bottom), three-domain cycle case}
\label{fig_bilin_eqlin_sym}
\end{figure}

According to the previous discussion, both two-domain and three-domain limit cycles will be asymmetric for $\alpha_p > 0$. In order to apply equivalent \revA{linearisation~\cite{krylov} (also known as the describing function or Krylov and Bogoliubov method)} we must look for limit cycles of the form
\begin{equation}
\alpha(t)=A\sin\omega t+\alpha_0
\label{alphadefine}
\end{equation}
where $A$ is the amplitude, $\omega$ is the frequency and $\alpha_0$ is the centre of the limit cycle. Three types of limit cycle can occur:
\begin{itemize}
\item Three-domain cycles, spanning $S_1$, $S_2$ and $S_3$
\item Two-domain cycles spanning $S_1$ and $S_2$
\item Two-domain cycle spanning $S_1$ and $S_3$
\end{itemize}
In all cases, the nonlinear restoring moment of equation~\ref{eq_freeplay} is approximated as a Fourier series of the form
\begin{equation}
M_{\alpha}(\alpha)=a_0+a_1\cos\omega t+b_1 \sin \omega t
\label{Malpha}
\end{equation}
where
\begin{eqnarray}
a_0 & = & \frac{\omega }{2\pi}\int_{-\pi/\omega }^{\pi/\omega }M(A \sin \omega t+\alpha_0) \rd t \nonumber \\
a_1 & = & \frac{\omega }{\pi}\int_{-\pi/\omega }^{\pi/\omega }M(A \sin\omega  t+\alpha_0) \cos \omega t \rd t \label{eq_Fourier} \\
b_1 & = & \frac{\omega }{\pi}\int_{-\pi/\omega }^{\pi/\omega }M(A \sin \omega t+\alpha_0) \sin \omega t \rd t \nonumber 
\end{eqnarray}
For a three-domain cycle, $A$ must be large enough that $\alpha(t)$ spans all three domains. Figure~\ref{fig_bilin_eqlin_sym} plots the pitch variation in time and the corresponding freeplay load for the case $\alpha_0=0.3$, $\delta=0.5$, $A=1$ and $\omega=1$. The integrals of equation~\ref{eq_Fourier} must be carried out in the intervals \revA{$[-\pi , t_1]$, $[t_1 , t_2]$, $[t_2 , t_3]$, $[t_3 , t_4]$ and $[t_4 , \pi]$}. These time instances are given by
$t_1=-\pi+\sin^{-1}(\delta+\alpha_0)/A$, $t_2=-\sin^{-1}(\delta+\alpha_0)/A$, $t_3=\sin^{-1}(\delta-\alpha_0)/A$, $t_4=\pi-\sin^{-1}(\delta-\alpha_0)/A$.
\LLL
\LLL
As an example, after defining 
\begin{eqnarray}
\sigma_1 & = & \sin^{-1}\left( \frac{\delta-\alpha_0}{A}\right) \\
\sigma_2 & = & \sin^{-1}\left( \frac{\delta+\alpha_0}{A}\right)  \label{sigma_12}
\end{eqnarray}
the equation for $b_1$ becomes
\begin{equation}
b_1 = \frac{1 }{\pi}\int_{-\pi+\sigma_2}^{-\sigma_2} K_{\alpha}(A\sin t+\delta)  \sin  t  \rd t  + \frac{1 }{\pi}\int_{\sigma_1}^{\pi-\sigma_1} K_{\alpha}(A\sin t-\delta)  \sin  t  \rd t  
\end{equation}
After working out the integrals and repeating for $a_0$ and $a_1$, equations~\ref{eq_Fourier} become
\begin{eqnarray}
a_0 & = & \frac{K_{\alpha}}{\pi}(\pi  \alpha_0 -  \alpha_0 (\sigma_1 + \sigma_2)  +  \delta (\sigma_1 - \sigma_2) + A  (\cos \sigma_1  - \cos \sigma_2 ))
 \nonumber \\
a_1 & = & 0 \\
b_1 & = &  \frac{A K_{\alpha}}{2\pi} \left(2 \pi  - ( \sin 2 \sigma_1 +  \sin 2 \sigma_2 )   - 2  (\sigma_1 + \sigma_2) \right)  \nonumber
\end{eqnarray}
\LLL
Looking back at equation~\ref{alphadefine}, it can be re-arranged as
\begin{equation}
\sin\omega t=\frac{\alpha-\alpha_0}{A}
\end{equation}
so that, after setting $a_1=0$, equation~\ref{Malpha} becomes
\begin{equation}
M_{\alpha}(\alpha)=a_0+b_1 \frac{\alpha-\alpha_0}{A}=a_0- \frac{b_1\alpha_0}{A} +\frac{b_1}{A}\alpha
\label{Mequiv}
\end{equation}
In this expression, there is a constant term and a term proportional to $\alpha$, i.e. a linear stiffness term. An equivalent linear stiffness can be defined as $K_{eq}=b_1/A$, or
\begin{equation}
K_{eq}=\frac{ K_{\alpha}}{2\pi} \left(2 \pi  - ( \sin 2 \sigma_1 +  \sin 2 \sigma_2 )   - 2  (\sigma_1 + \sigma_2) \right)
\label{Keq3}
\end{equation}
\revA{Replacing the nonlinear function in equations~\ref{eq_dof3_23}} by the equivalent linear function of equation~\ref{Malpha}, the complete equivalent linear system for three-domain cycles is obtained as
\begin{equation}
\dot{\mathbf{x}}=\mathbf{Q}_{eq}\mathbf{x}+\mathbf{q}_n \left( a_0 -K_{eq} \alpha_0\right)+\mathbf{q}_p a_p
\label{eom_eqlin_asym}
\end{equation}
for different values of $A$ and $\alpha_0$, where $\mathbf{Q}_{eq} \mathbf{x}=\mathbf{Q}_1 \mathbf{x} + \mathbf{q}_n K_{eq}(A,\alpha_0) \alpha$. Note that the fixed point of the equivalent linearised system is given by
\begin{equation}
\mathbf{x}_{eq}=-\mathbf{Q}_{eq}^{-1} \left( \mathbf{q}_n \left( a_0 -K_{eq} \alpha_0\right) +\mathbf{q}_p a_p \right)
\label{xf_eq}
\end{equation}
The equivalent linearised system can only exist if both $\sigma_1$ and $\sigma_2$ are real. This means that
\begin{equation}
\begin{array}{cc}
 |\delta-\alpha_0| & \leq  A  \\
 |\delta+\alpha_0| & \leq  A 
\end{array}
\end{equation}
simultaneously. 
 \revA{If $\alpha_p$ is not equal to zero then $\alpha_0$ is also not equal to zero and is it not possible to obtain a symmetric response.} Under these circumstances, three-domain limit cycles with amplitude $A=\delta$ \revA{cannot exist} because they would violate one of the \revA{conditions of existence}. 
\LLL
The case $K_{eq}=0$ reflects the bifurcation condition for the symmetric system, i.e. $\alpha_p=0$. A limit cycle with the lowest possible amplitude $A=\delta$ appears at $U_{F_1}$, the flutter speed of the underlying linear system. However, for an asymmetric system this limit cycle cannot occur, as explained \revA{earlier}. Therefore, $K_{eq}=0$ is not the bifurcation condition for asymmetric systems and limit cycles will start appearing at airspeeds higher than $U_{F_1}$.
\LLL
The equivalent linearisation problem consists in determining the values of $A$ and $\alpha_0$ that result in periodic \revA{solutions} at each airspeed value of interest. The solution process is identical to the one developed in~\cite{Dimitriadis15}. First, the flutter speed of the equivalent linearized system, $U_{F_{eq}}(K_{eq})$ is calculated for all values of $K_{eq}$ from 0 to $K_{\alpha}$. Then, for each $U_{F_{eq}}(K_{eq})$ we calculate the values of $A$ and $\alpha_0$ that give the correct value of the equivalent linearised stiffness when substituted into equation~\ref{Keq3} and for which $\alpha_0$ is equal to the pitch component of the fixed point of the equivalent linearised system. The nonlinear algebraic system
\begin{eqnarray}
K_{eq}- \frac{ K_{\alpha}}{2\pi} \left(2 \pi  - ( \sin 2 \sigma_1 +  \sin 2 \sigma_2 )   - 2  (\sigma_1 + \sigma_2) \right)  & = & 0 \nonumber \\
\alpha_{F_{eq}}-\alpha_0 & = & 0 \label{eq_solve3domain} 
\end{eqnarray}
can be set up and solved using Newton iterations, where $\alpha_{F_{eq}}$ is the pitch component of $\mathbf{x}_{eq}$ from equation~\ref{xf_eq}. The starting point of the limit cycle branch is $K_{eq}=0$, $U=U_{F_1}$. Initial  guesses are chosen as $A=\delta$ and $\alpha_0=0$. 

Once a converged limit cycle is obtained, its stability can be investigated by evaluating a new equivalent linearized system at the same airspeed but with slightly higher amplitude, i.e. $A+\delta A$, where $\delta A << 1$.  We calculate the new value of $\alpha_0$ that corresponds to this amplitude and the resulting $K_{eq}$. Finally, we set up the new equivalent linearised system using equations~\ref{eom_eqlin_asym}. If all of its eigenvalues have negative real part, then the limit cycle at $A$ is stable. If any of the eigenvalues have a positive real part then the limit cycle is unstable.
\LLL
Two-domain limit cycles can be approximated using the same equivalent linearisation scheme. The only difference lies in the values of $K_{eq}$ and $\alpha_0$, which are different for two-domain cycles. \revA{Applying the Fourier series procedure} to two-domain cycles  gives the following expressions for $a_0$, $a_1$ and $b_1$~\cite{Dimitriadis15}
\begin{eqnarray}
a_0 & = & \frac{K_{2}\alpha_0}{2}-\frac{K_{2}\delta}{2}+\frac{AK_{2}}{\pi}\left(\sigma_{1}\sin\sigma_{1}+ \cos\sigma_{1} \right) 
\nonumber \\
a_1 & = & 0 \label{eq_a0_bilin_asym} \\
b_1  & = & A\frac{K_{2}}{2}-A\frac{K_{2}}{2\pi}\left(2\sigma_{1}+\sin 2\sigma_{1} \right)  \nonumber
\end{eqnarray}
so that $K_{eq}$ \revA{becomes}
\begin{equation}
K_{eq}  = \frac{K_{2}}{2}-\frac{K_{2}}{2\pi}\left(2\sigma_{1}+\sin 2\sigma_{1} \right) 
\label{eq_Keq_bilin_asym}
\end{equation}
Two-domain limit cycles can exist as long as $ |\delta-\alpha_0| \leq  A$, i.e. $\sigma_1$ is real. Furthermore, they must only span two domains so that, 
\begin{eqnarray}
\mathrm{if \hspace{2ex}} \alpha_0+A \geq \delta &  \mathrm{then \hspace{2ex}} & \alpha_0-A \geq -\delta \nonumber \\ 
\mathrm{if \hspace{2ex}} \alpha_0-A \leq -\delta &  \mathrm{then \hspace{2ex}} & \alpha_0+A \leq \delta \label{twodomain_conds}
\end{eqnarray}
This means that neither the upper nor the lower bound of the cycle can cross a freeplay boundary. For example, a two-domain cycle spanning $S_1$ and $S_2$ will disappear if either of the two bounds crosses $+\delta$. Note that $\sigma_1$ takes values between $-\pi/2 \leq \sigma_1 \leq \pi/2$, while $\delta -\alpha_0$ spans
$-A \leq \delta -\alpha_0 \leq A$. 

For the right limit $\sigma_1=\pi/2$, $ \delta -\alpha_0=A$, substituting into equations~\ref{eq_a0_bilin_asym} and~\ref{eq_Keq_bilin_asym} yields
\begin{equation}
a_0=0 \mbox{, \hspace{2ex}} K_{eq}=0 \mbox{, \hspace{2ex}}
\end{equation}
while for the left limit $\sigma_1=-\pi/2$, $ \delta -\alpha_0=-A$ we obtain
\begin{equation}
a_0=K_{\alpha}(\alpha_0+\delta) \mbox{, \hspace{2ex}} K_{eq}=K_{\alpha}
\end{equation}
In other words, two-domain limit cycles appear when the equivalent stiffness is equal to the stiffness of the underlying linear system and disappear when $K_{eq}$ is equal to the stiffness of the overlying linear system. Again, these conclusions are only true if $\alpha_p=0$\revA{;} in the presence of aerodynamic preload the $K_{eq}=0$ and $K_{eq}=K_{\alpha}$ appearance and disappearance bounds are modified.

\section{\revA{Comparison of the mathematical predictions and experimental observations}} 
\begin{figure}[ht]
  \begin{center}
  \subfigure[Freeplay = 1 deg and $\theta = 0.5 $ deg]{\label{MvsEA01} \includegraphics[height=.35\textwidth]{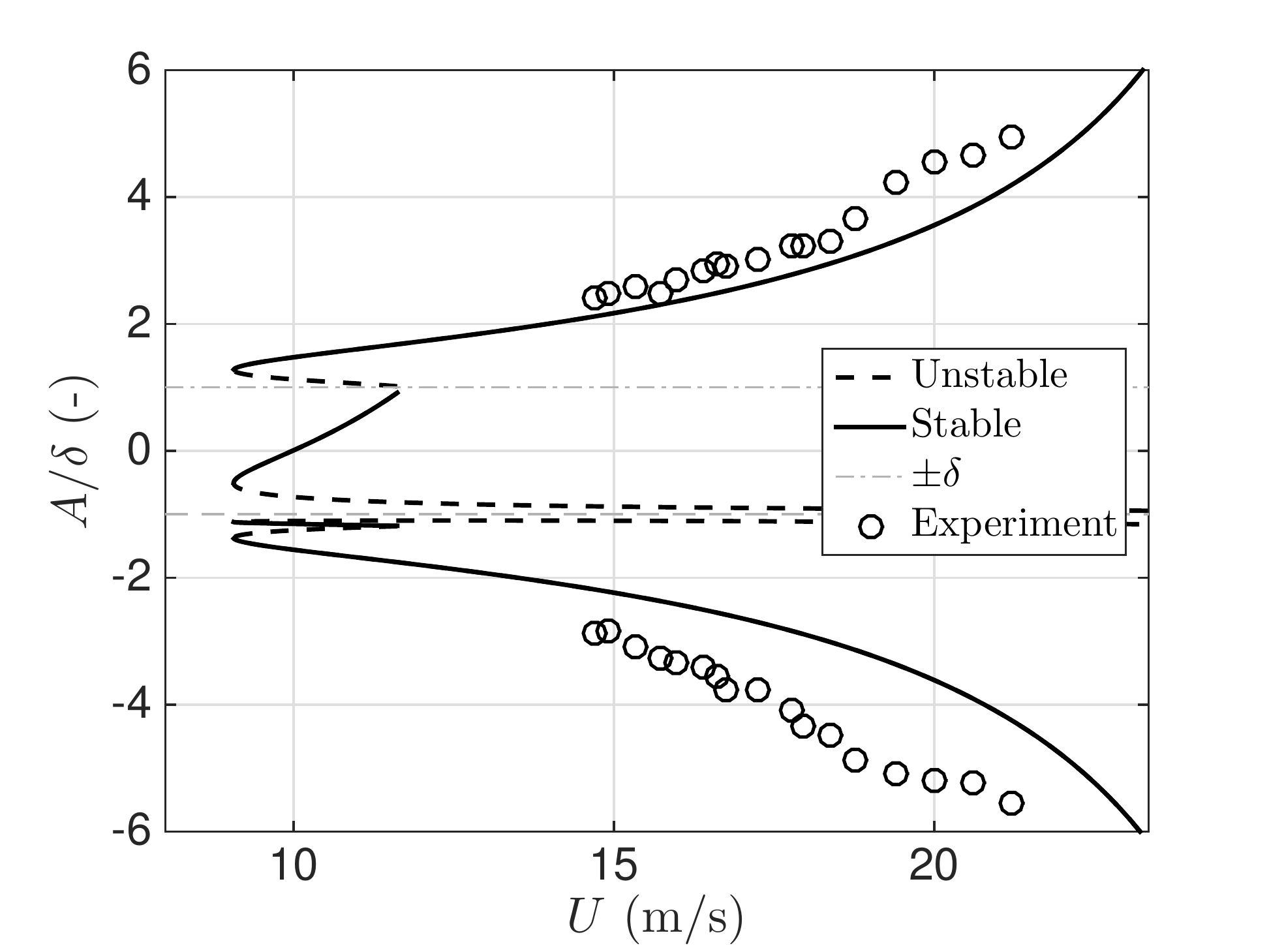}}
  \subfigure[Freeplay = 2 deg and $\theta = 0.7 $ deg]{\label{MvsEA02} \includegraphics[height=.35\textwidth]{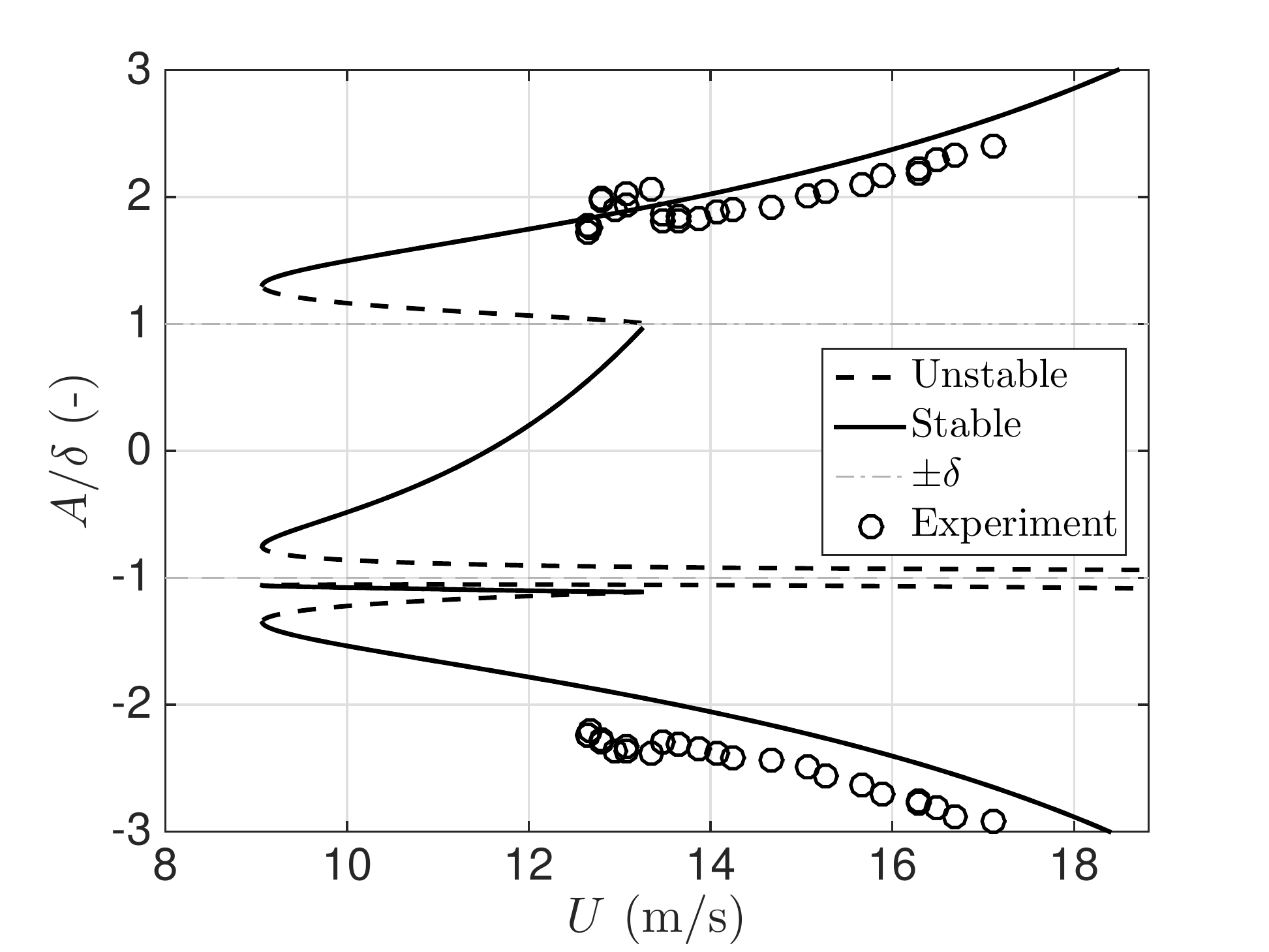}}
  \subfigure[Freeplay = 3 deg and $\theta = 1.5 $ deg]{\label{MvsEA03} \includegraphics[height=.35\textwidth]{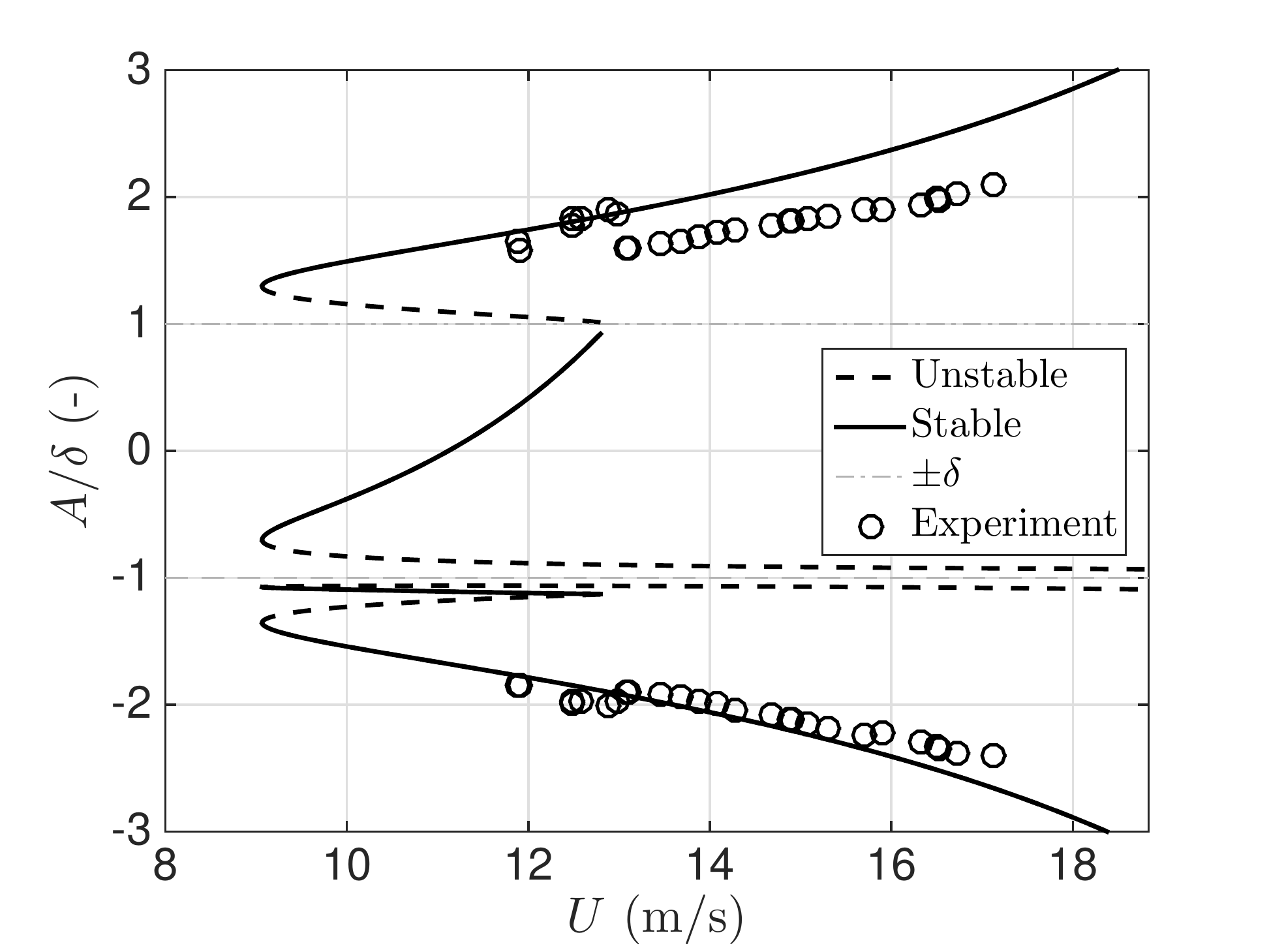}}
  \subfigure[Freeplay = 8 deg and $\theta = 3 $ deg]{\label{MvsEA04} \includegraphics[height=.35\textwidth]{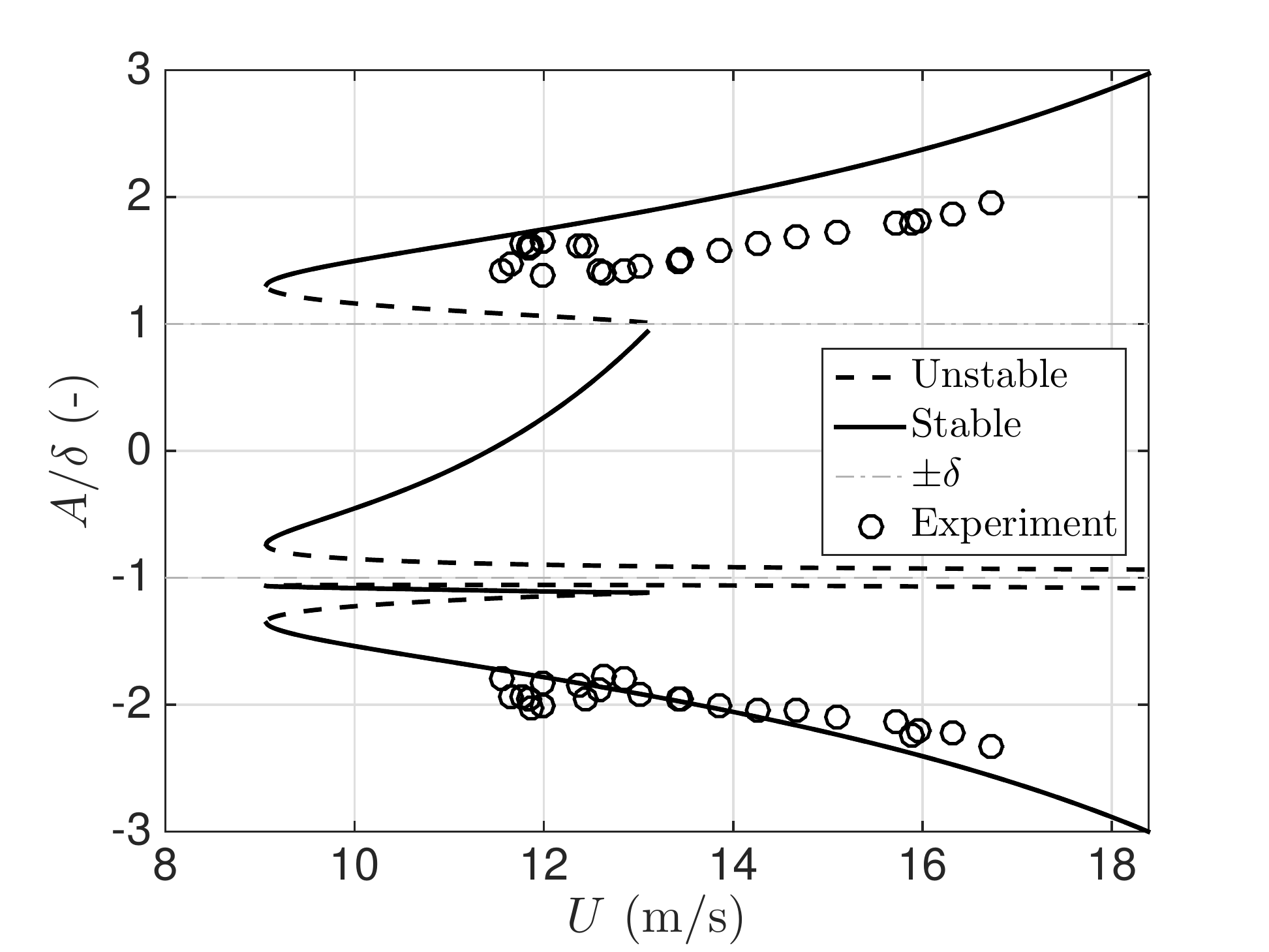}}
  \end{center}
  \caption{Pitch amplitude bifurcation diagram of the system with a preload angle of 0 deg}
\label{mod_vs_exp_ampl0}
\end{figure}

\begin{figure}[ht]
  \begin{center}
  \subfigure[Freeplay = 1 deg and $\theta = 0.5 $ deg]{\label{MvsEF01} \includegraphics[height=.35\textwidth]{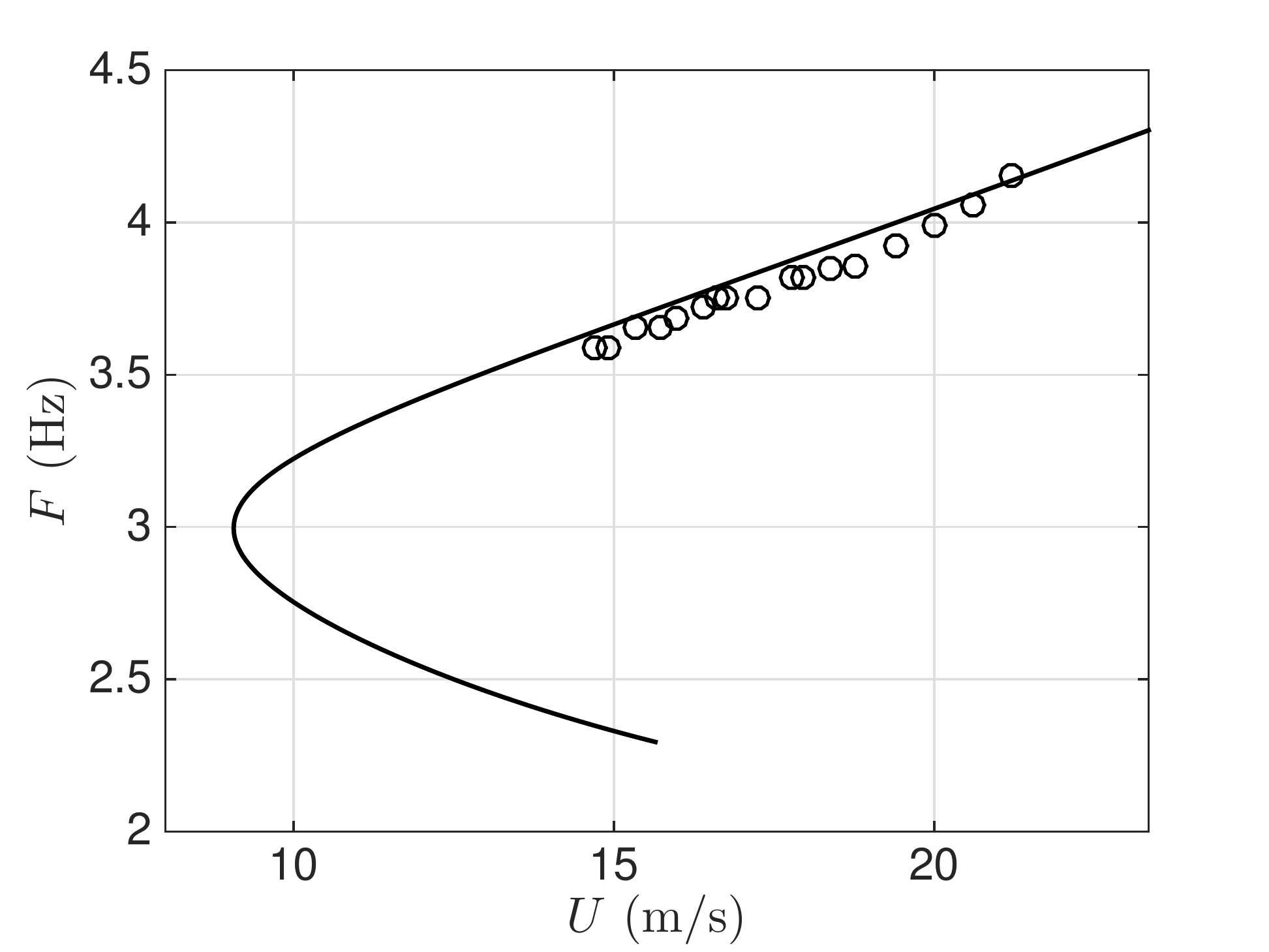}}
  \subfigure[Freeplay = 2 deg and $\theta = 0.7 $ deg]{\label{MvsEF02} \includegraphics[height=.35\textwidth]{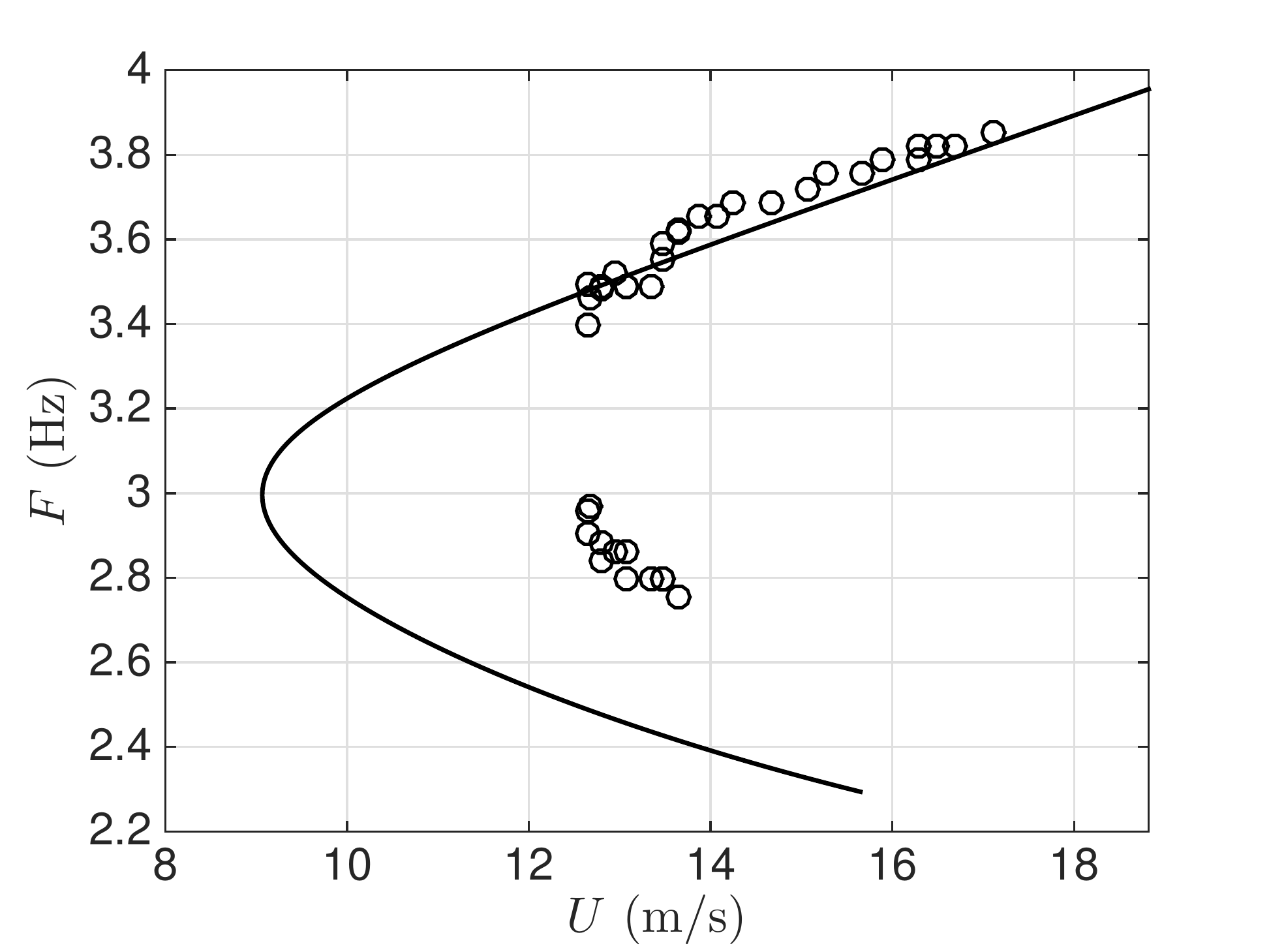}}
  \subfigure[Freeplay = 3 deg and $\theta = 1.5 $ deg]{\label{MvsEF03} \includegraphics[height=.35\textwidth]{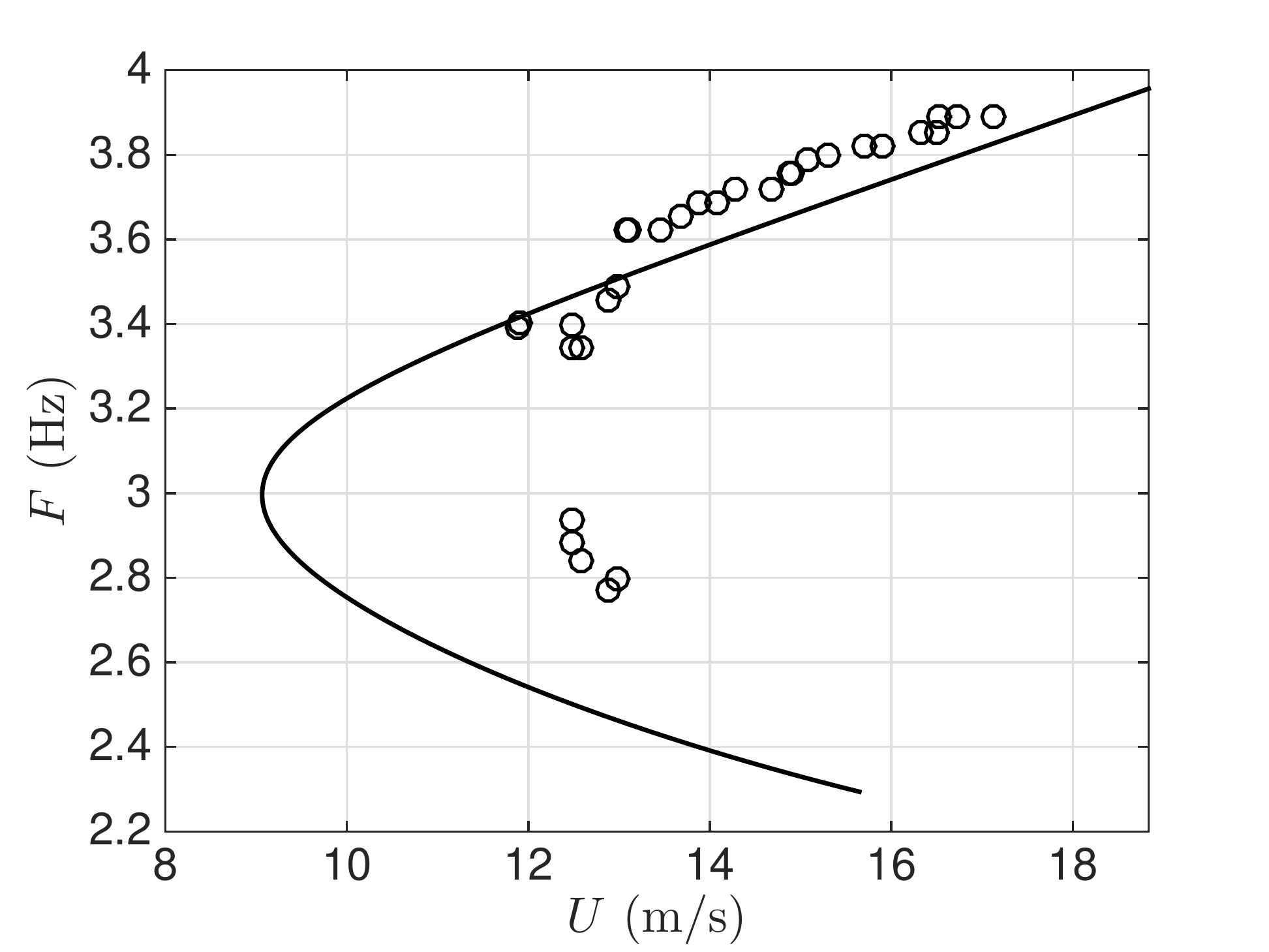}}
  \subfigure[Freeplay = 8 deg and $\theta = 3 $ deg]{\label{MvsEF04} \includegraphics[height=.35\textwidth]{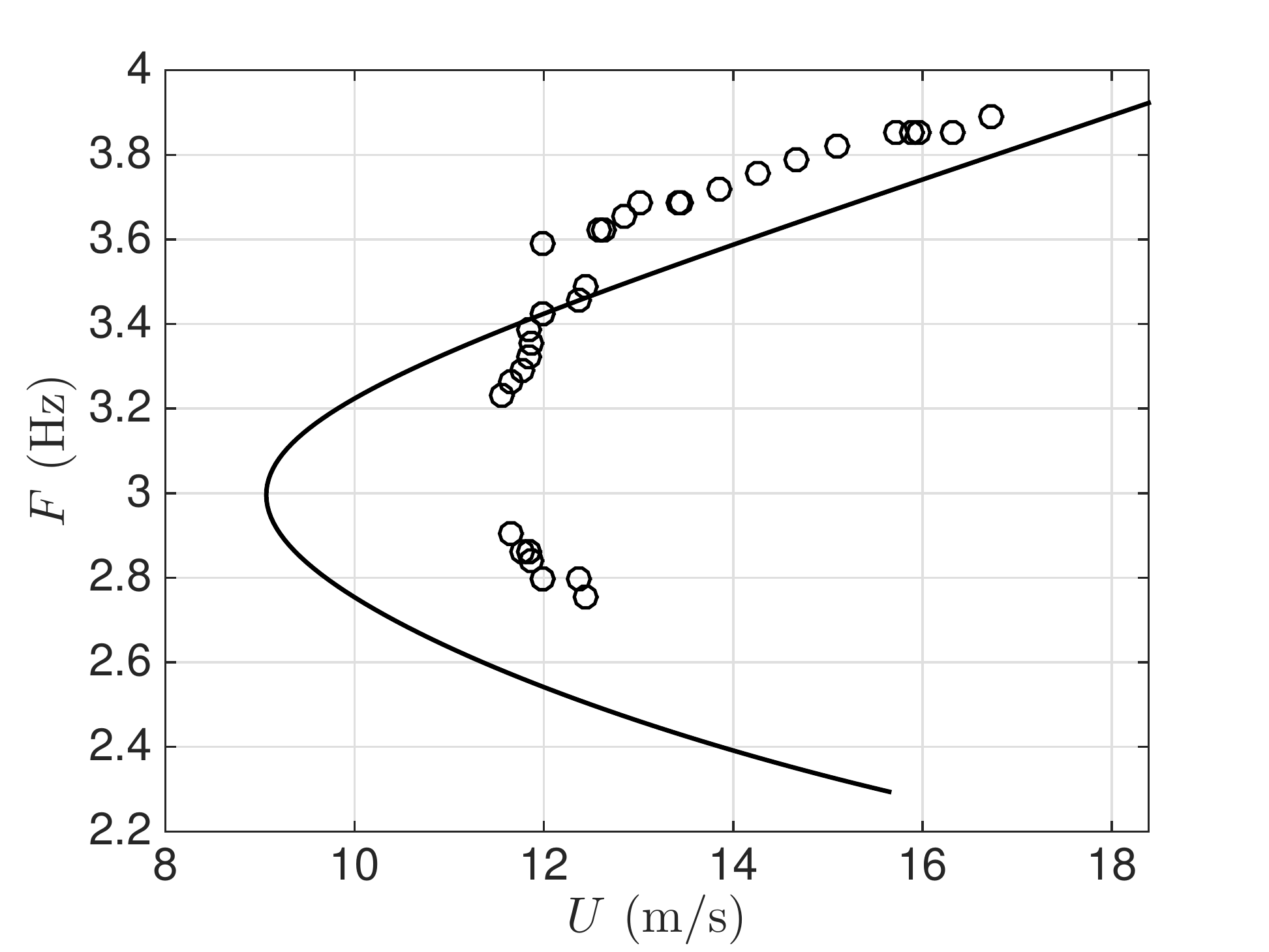}}
  \end{center}
  \caption{Frequency bifurcation diagram of the system with a preload angle of 0 deg}
\label{mod_vs_exp_freq0}
\end{figure}

\begin{figure}[ht]
  \begin{center}
  \subfigure[Freeplay = 1 deg and $\theta = 0 $ deg]{\label{MvsEA51} \includegraphics[height=.35\textwidth]{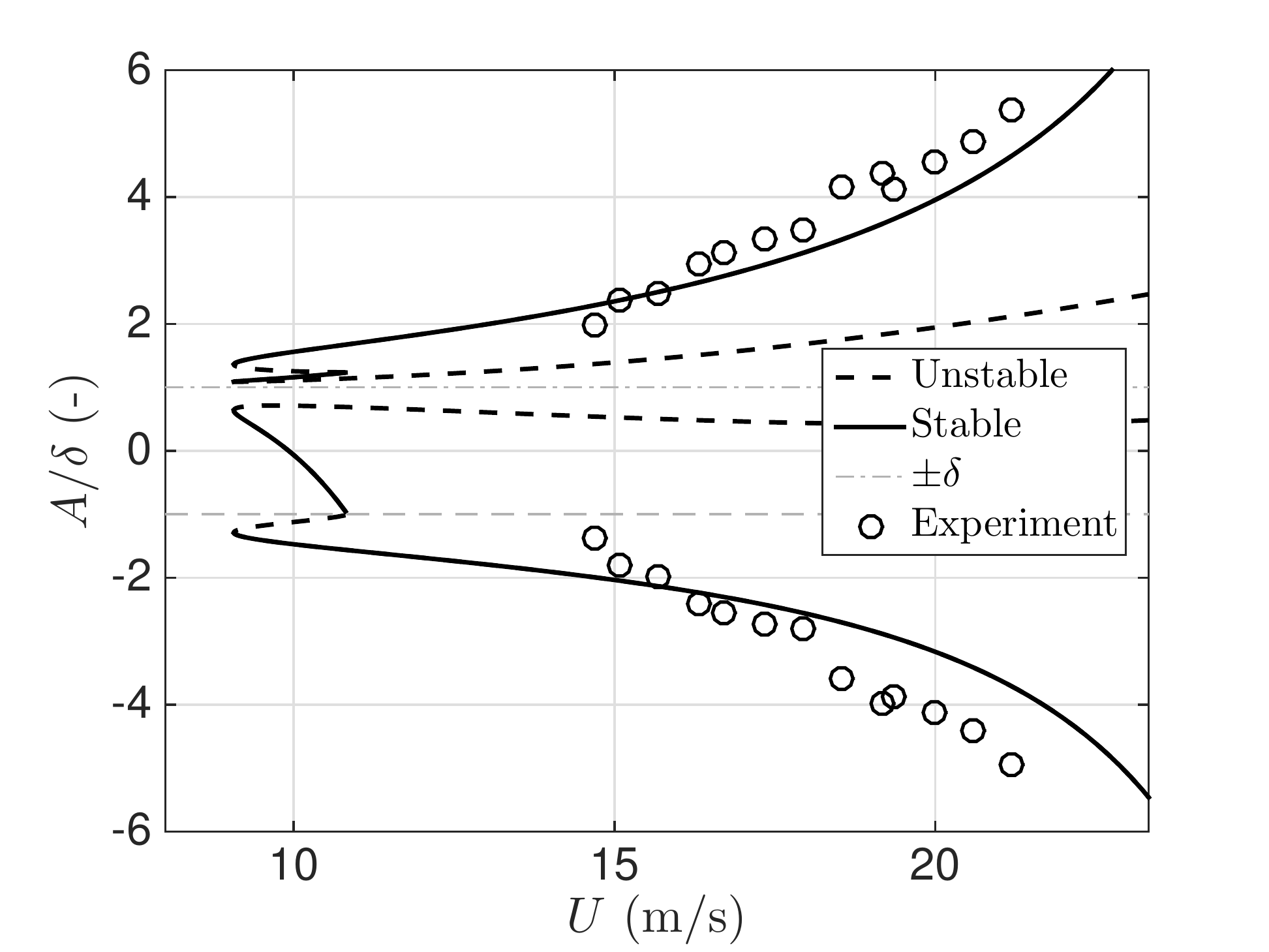}}
  \subfigure[Freeplay = 2 deg and $\theta = 0.5 $ deg]{\label{MvsEA52} \includegraphics[height=.35\textwidth]{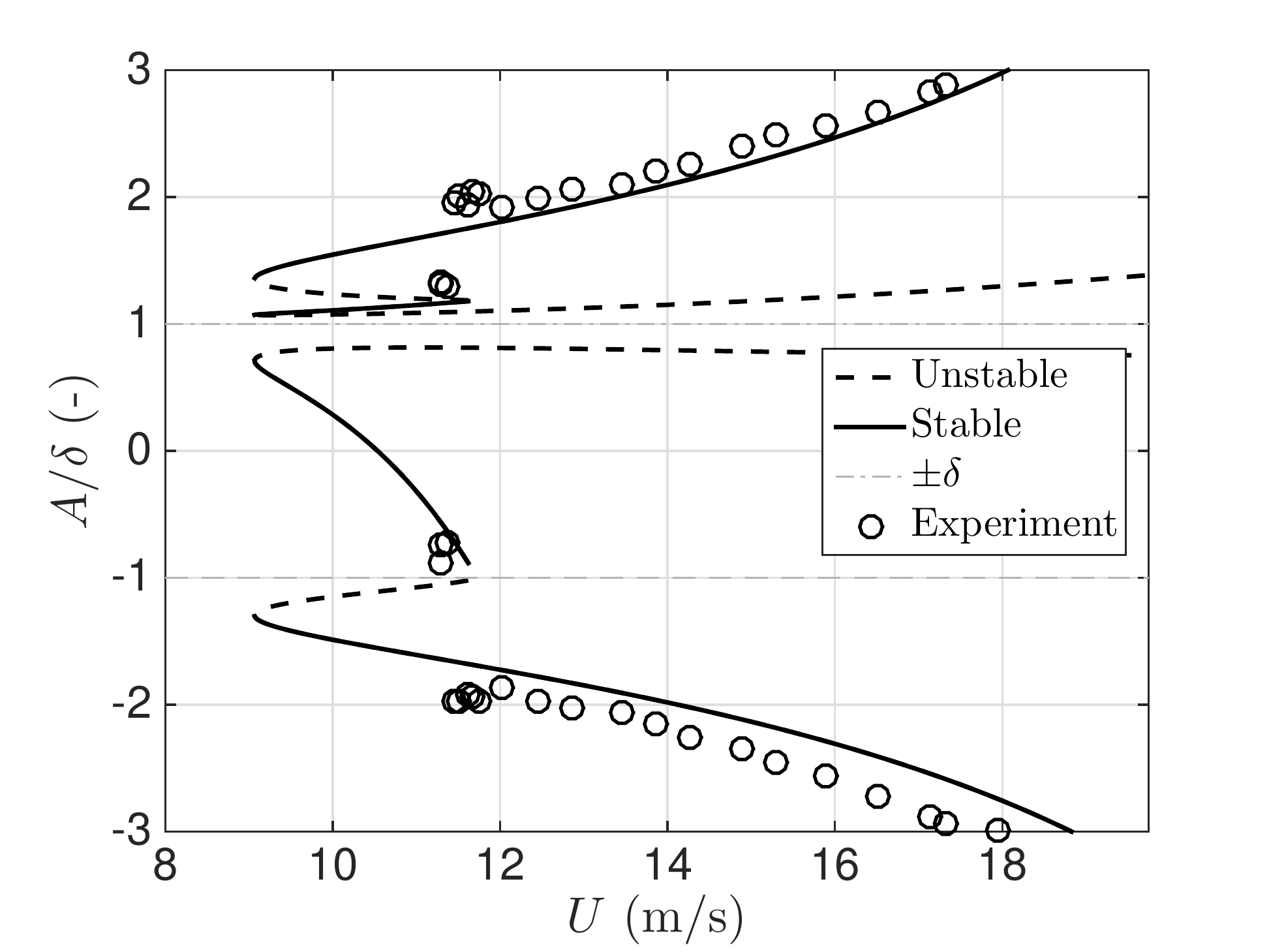}}
  \subfigure[Freeplay = 3 deg and $\theta = 1.25 $ deg]{\label{MvsEA53} \includegraphics[height=.35\textwidth]{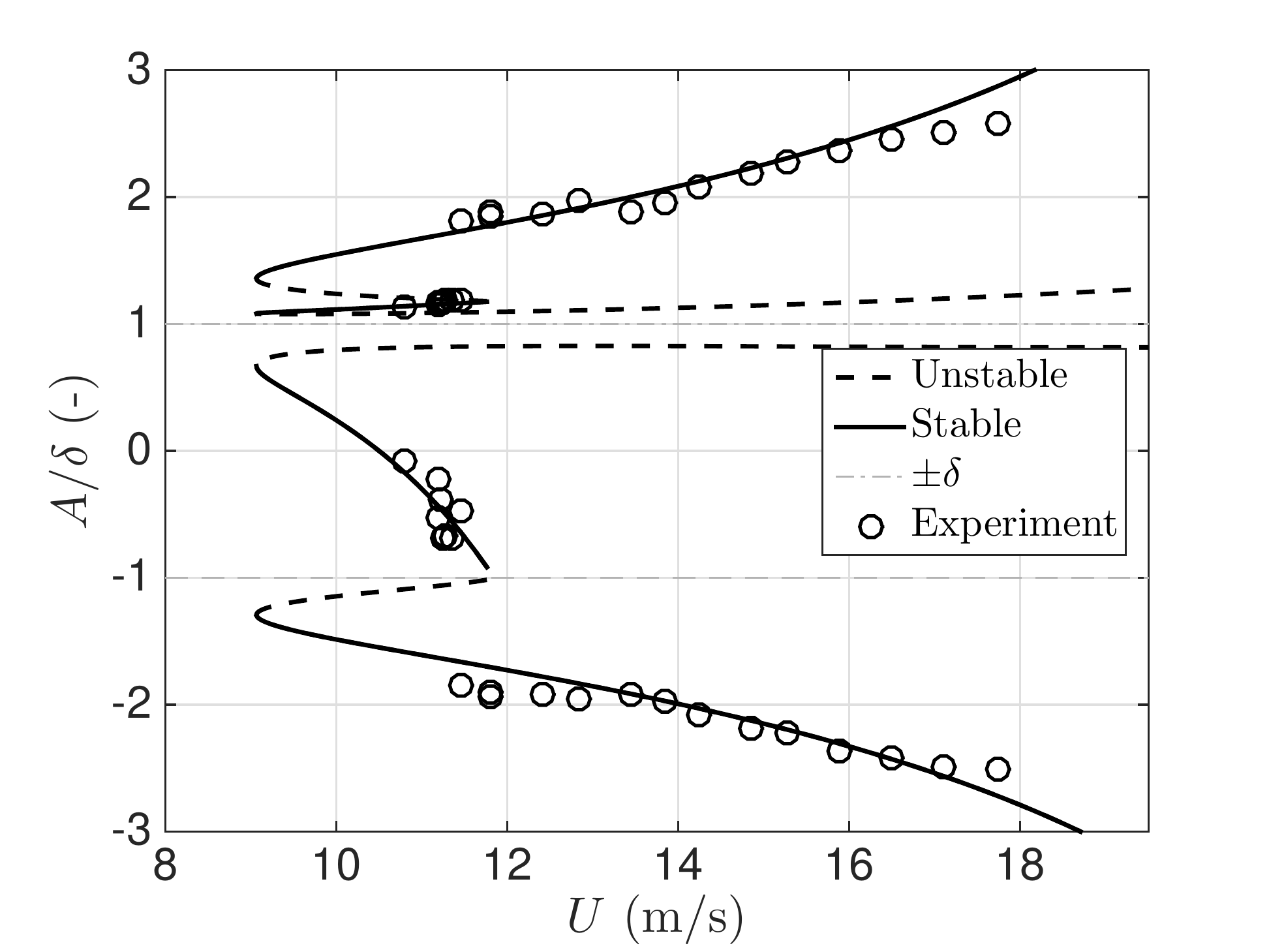}}
  \subfigure[Freeplay = 8 deg and $\theta = 3 $ deg]{\label{MvsEA54} \includegraphics[height=.35\textwidth]{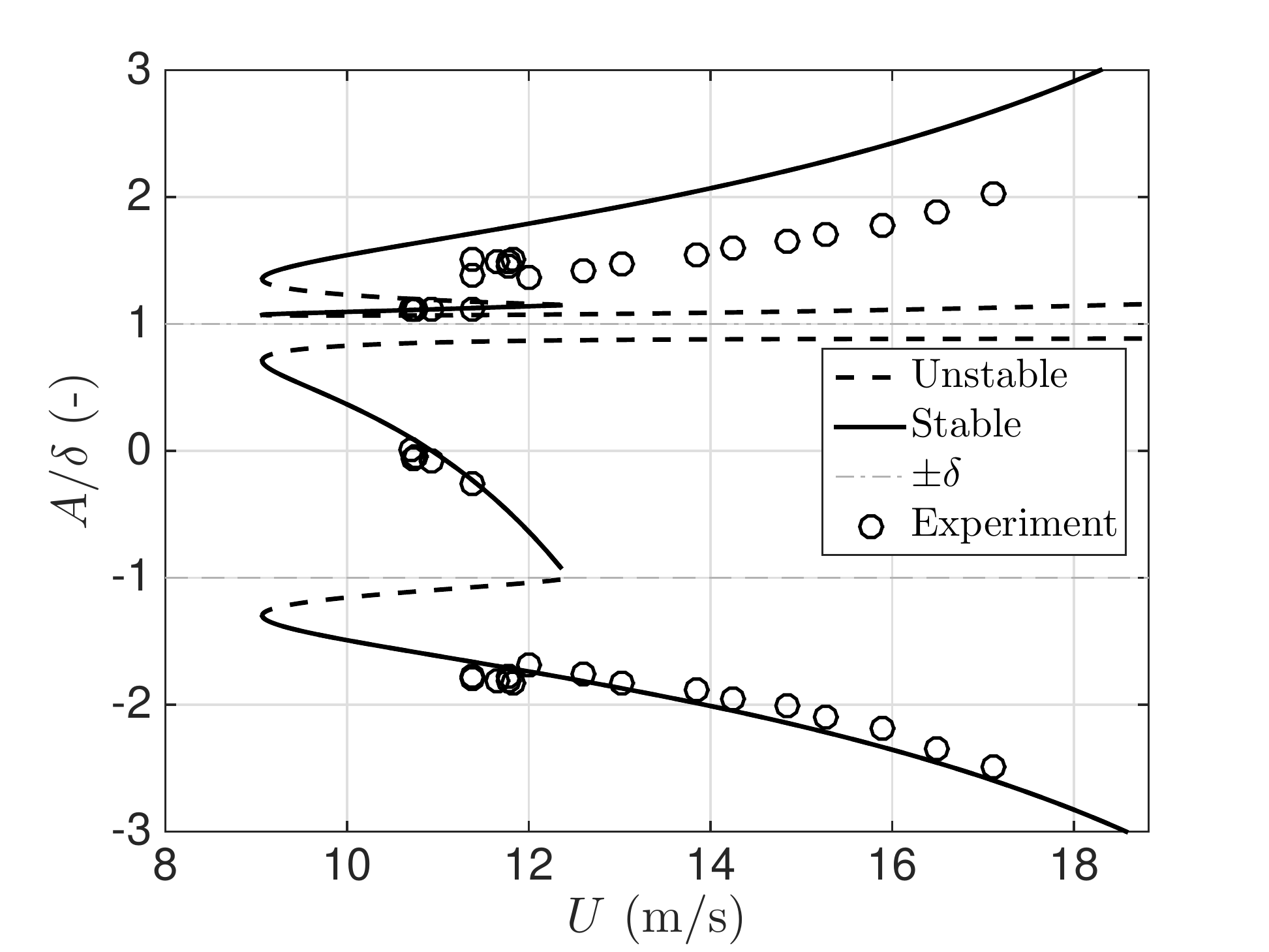}}
  \end{center}
  \caption{Pitch amplitude bifurcation diagram of the system with a preload angle of 5 degrees}
\label{mod_vs_exp_ampl5}
\end{figure}

\begin{figure}[ht]
  \begin{center}
  \subfigure[Freeplay = 1 deg and $\theta = 0 $ deg]{\label{MvsEF51} \includegraphics[height=.35\textwidth]{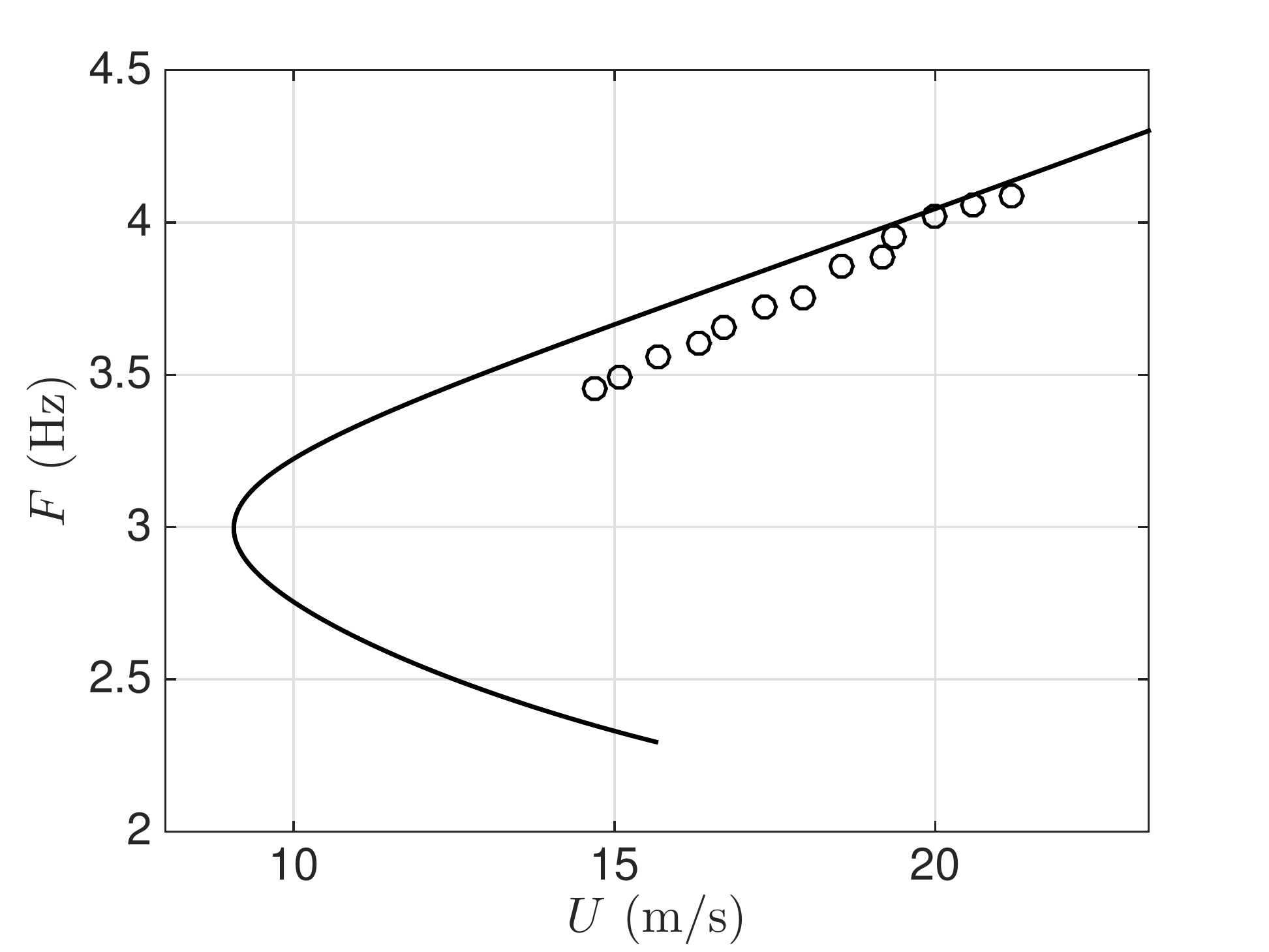}}
  \subfigure[Freeplay = 2 deg and $\theta = 0.5 $ deg]{\label{MvsEF52} \includegraphics[height=.35\textwidth]{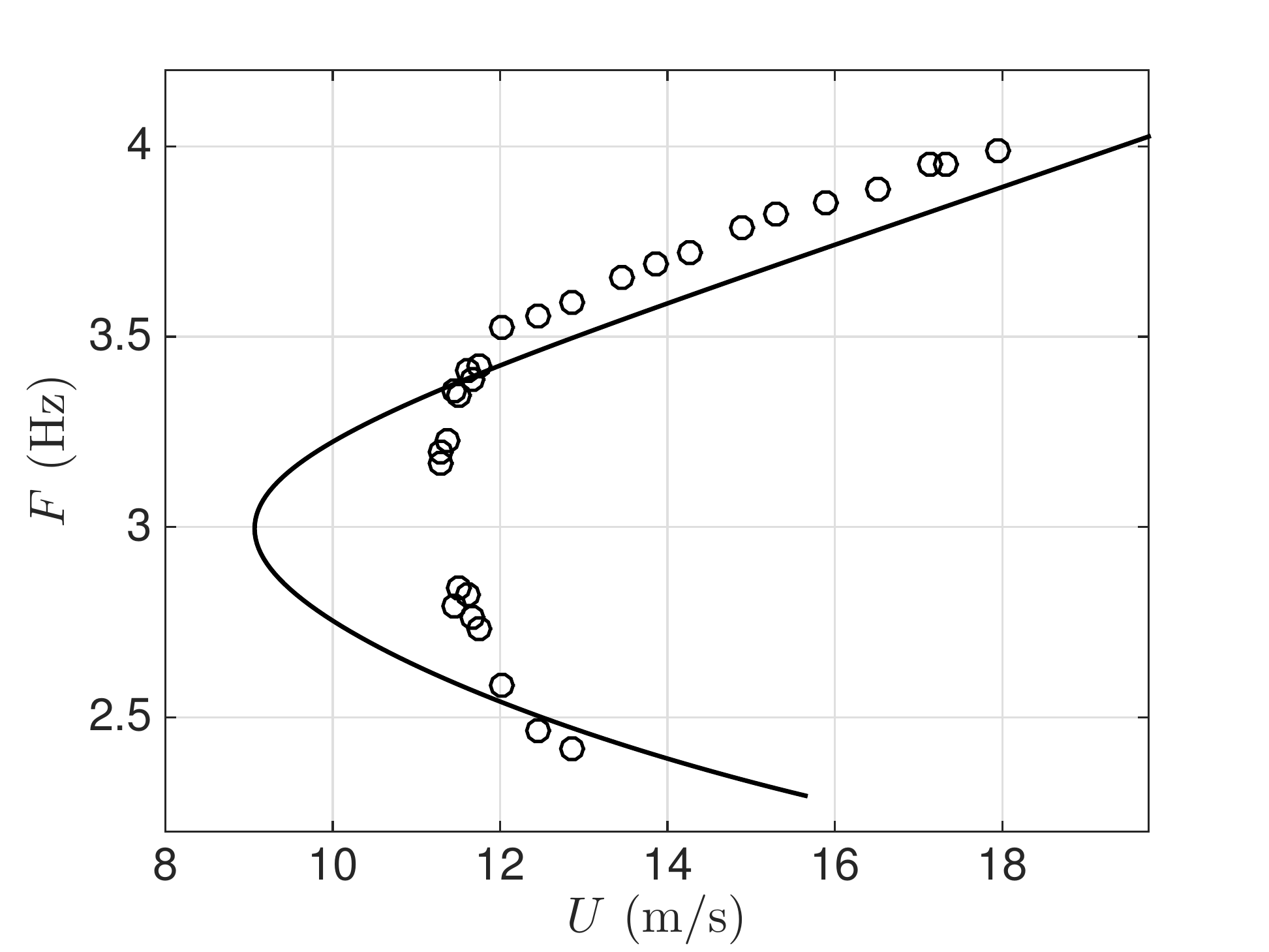}}
  \subfigure[Freeplay = 3 deg and $\theta = 1.25 $ deg]{\label{MvsEF53} \includegraphics[height=.35\textwidth]{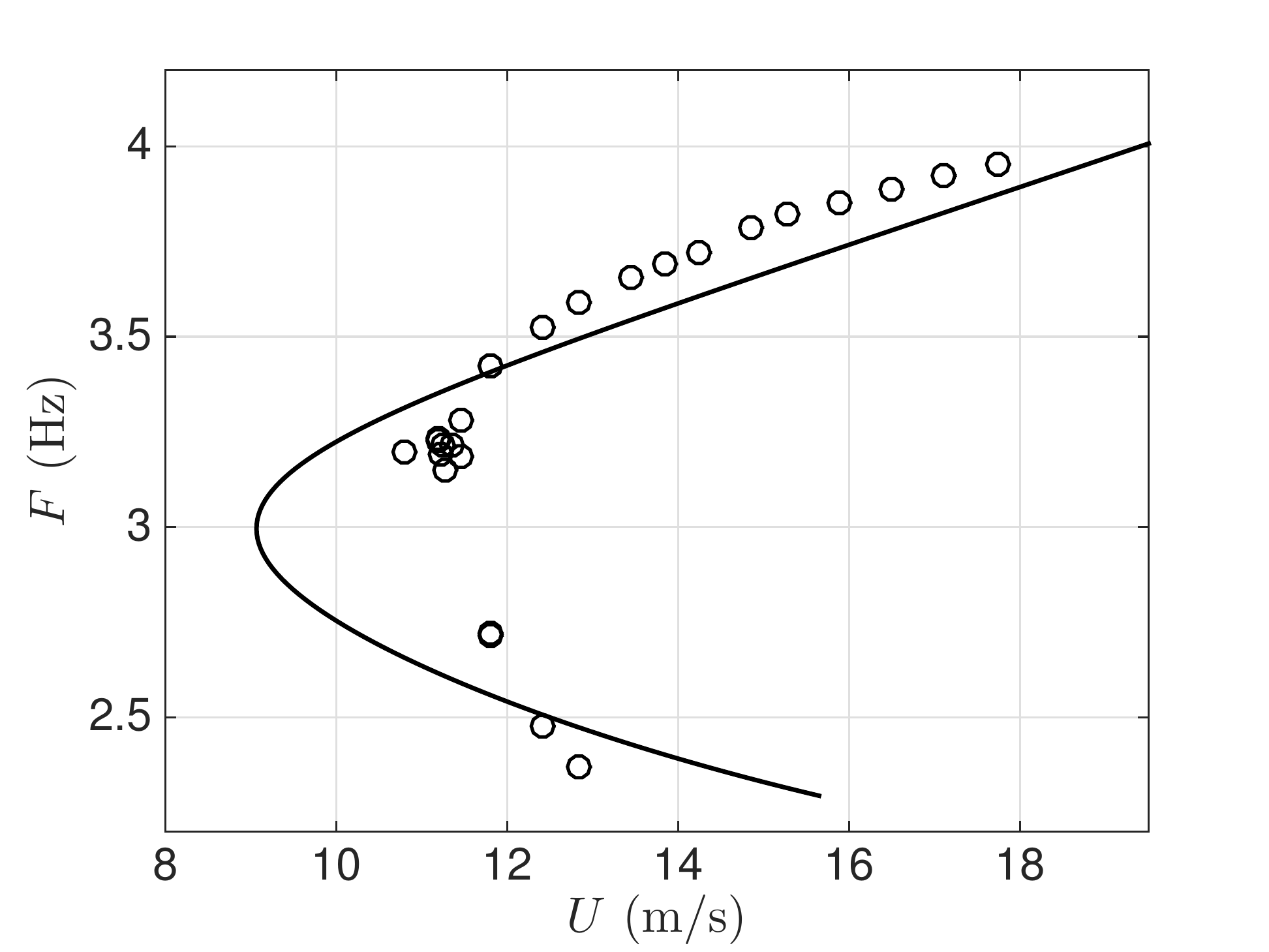}}
  \subfigure[Freeplay = 8 deg and $\theta = 3 $ deg]{\label{MvsEF54} \includegraphics[height=.35\textwidth]{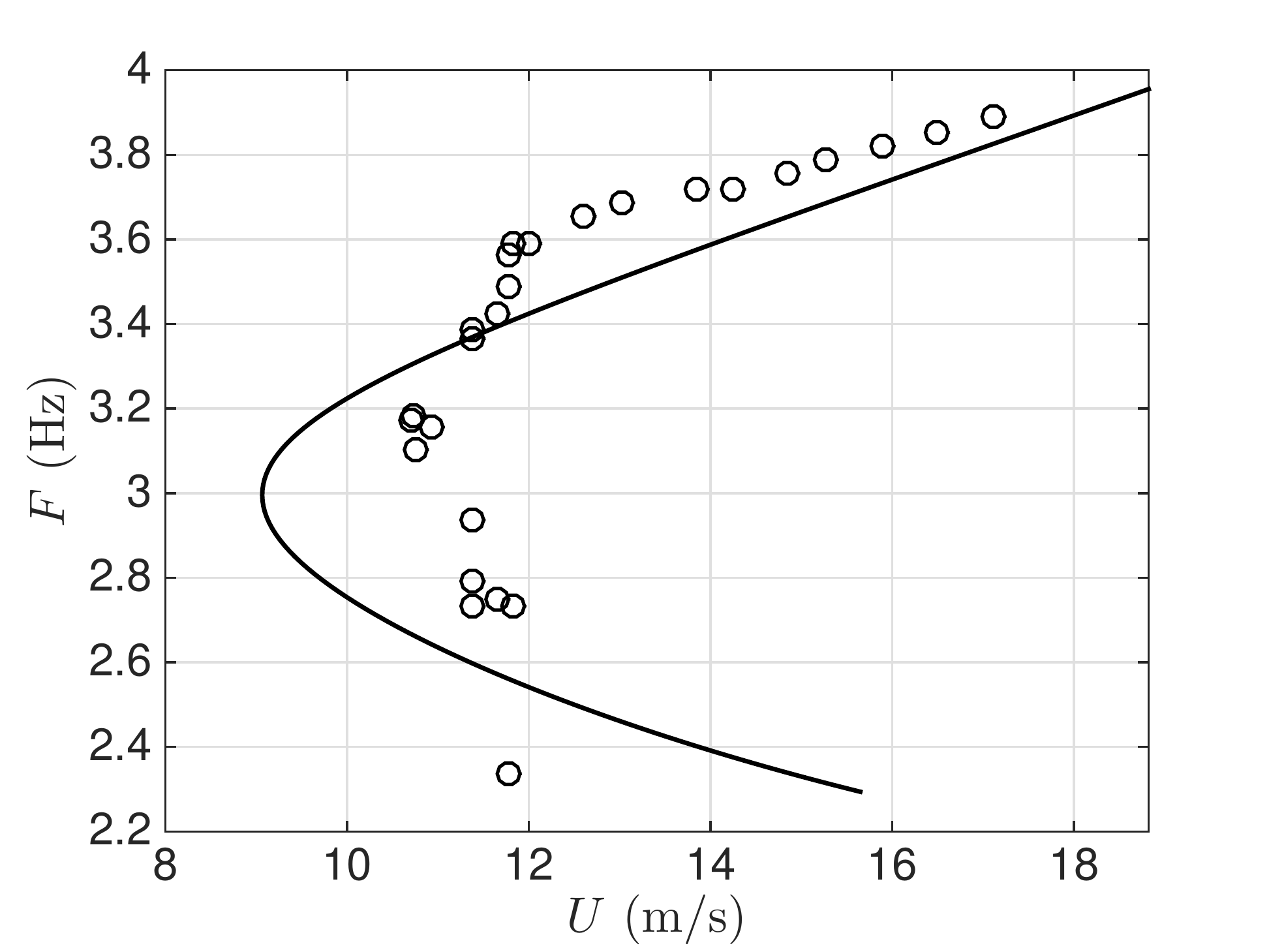}}
  \end{center}
  \caption{Frequency bifurcation diagram of the system with a preload angle of 5 degrees}
\label{mod_vs_exp_freq5}
\end{figure}

\begin{figure}[ht]
  \begin{center}
\includegraphics[height=.35\textwidth]{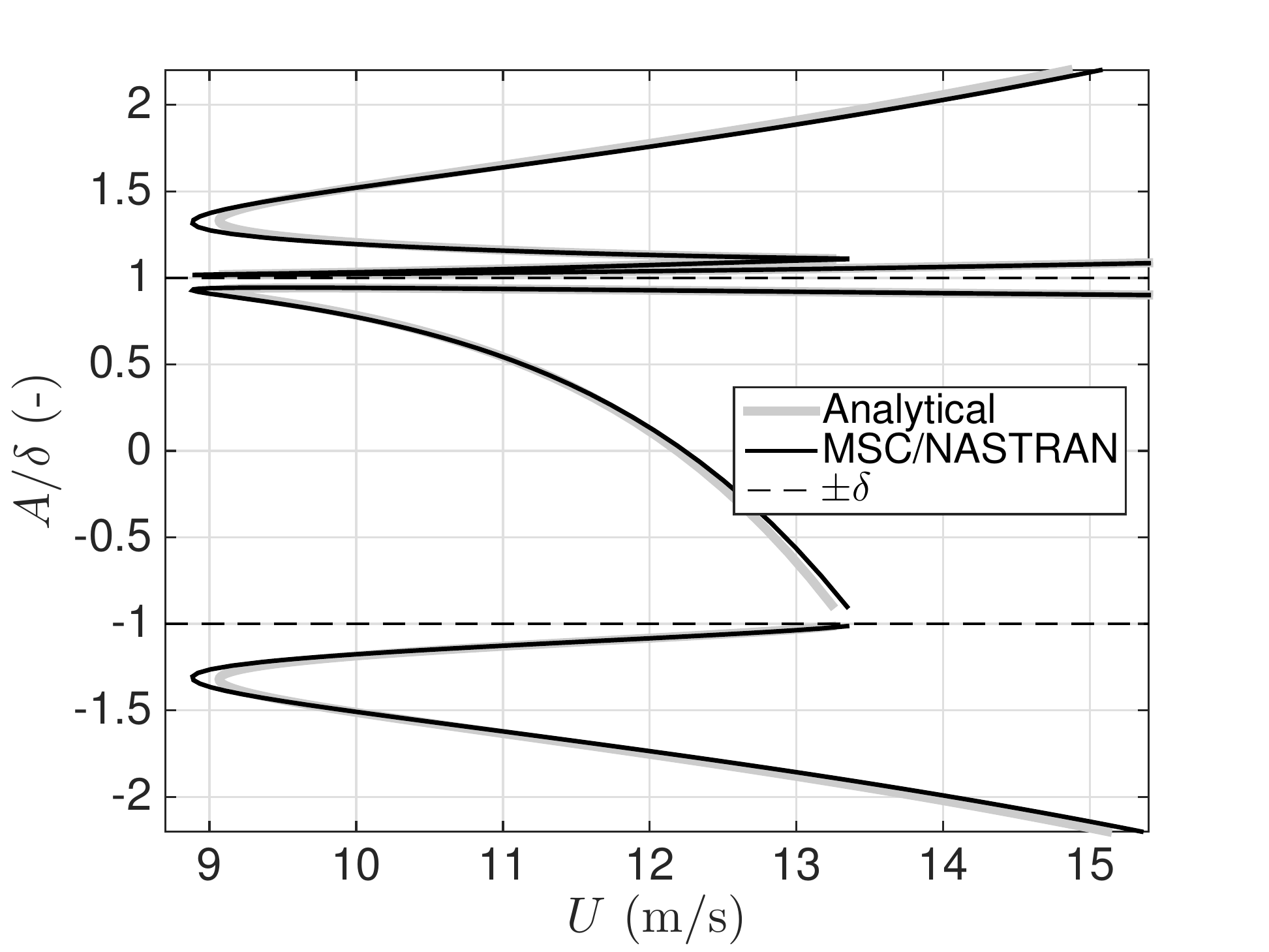}
  \end{center}
  \caption{Comparison of the analytical results to MSC/NASTRAN}
\label{msc_vs_math}
\end{figure}

A comparison of the theoretical and experimental limit cycle amplitude variation with airspeed for $\alpha_p = 0$ is plotted in figure \ref{mod_vs_exp_ampl0}. For all the \revA{freeplay gaps} considered, the mathematical model predicts two-domain LCOs, which were not observed in practice. In contrast, the amplitudes of the three-domain cycles \revA{are} predicted with satisfactory accuracy. In all four cases, the predicted two-domain limit cycle branch \revA{lies} in the airspeed range where quasi-periodic motion was observed in the experiment. That is consistent with the fact that the quasi-periodicity is due to the co-existence of nearby limit cycles at the same airspeed. In all four cases, the model predicts limit cycles at airspeeds lower than the experimental LCO onset speed. Again, it is believed that this disparity is due to the nonlinear damping present in the bearings of the experimental system, which suppressed small amplitude oscillations. 
\LLL
The frequency bifurcation diagram for the system with $\alpha_p = 0$ is depicted in figure \ref{mod_vs_exp_freq0}. The model accurately predicts the 
three-domain limit cycle frequency for all the freeplay values. However, the predicted two-domain frequencies (i.e. the part of the branch below 3 Hz) are slightly  lower than those observed experimentally.  
\LLL
Figure \ref{mod_vs_exp_ampl5} displays the experimental and mathematical pitch amplitude bifurcation diagrams of the system with an aerodynamic preload angle of 5 deg. The model estimates well both the two-domain and the three-domain LCO amplitudes for freeplay \revA{gaps} of 2 and 3 deg, slightly underestimates the amplitude when the freeplay \revA{gap is} 1 degree and slightly overestimates the top branch for 8 degree of freeplay \revA{gap}. In all four cases, the model predicts limit cycles at airspeeds lower than those observed in the experiment. These small amplitude vibrations are once again damped by the nonlinear friction in the bearings. This is especially noticeable in the 1 deg freeplay cases because the gap is so small that all LCOs \revA{have} low amplitude. It is important to note that, for freeplay \revA{gaps} of 2, 3 and 8 deg, the model predicts with satisfactory accuracy the amplitudes of both the two-domain and three-domain limit cycles.
\LLL
The observed and predicted \revA{frequency} variations with airspeed for $\alpha_p=5$ deg are compared in figure \ref{mod_vs_exp_freq5}. In this case, the frequencies of both the two-domain and three-domain limit cycles are predicted with satisfactory accuracy, although the lowest airspeed at which LCOs can occur is still under-predicted.
\LLL
\revA{The analytical and MSC/NASTRAN models gave nearly identical \revA{limit cycle amplitude and frequency predictions} for all test cases. For the sake of clarity, sample MSC/NASTRAN predictions are plotted in figure \ref{msc_vs_math} but only the estimates of the analytical model are shown in the other figures.} 
\section{Conclusions} 

This paper demonstrated mathematically and experimentally the co-existence of two-domain and three-domain limit cycles in aeroelastic systems with freeplay in pitch. The two-domain cycles have small amplitudes, occur at low airspeeds and can be quite irregular or quasi-periodic. The mathematical treatment predicts such cycles in cases with and without aerodynamic preload. However, the limit cycles occurring without preload have even smaller amplitudes and are even less periodic; as a consequence, they were observed intermittently in the experimental work. In contrast, two-domain cycles were clearly observed \revA{in cases} with preload, both experimentally and mathematically.

From a mathematical point of view, the two-domain and three-domain cycles can be seen as lying on the same limit cycle branch. The former are the result of a grazing bifurcation of the latter. When one of the bounds of a three-domain limit cycle grazes a discontinuity boundary and then enters the freeplay range, the cycle is transformed into a two-domain cycle and undergoes a fold bifurcation. The amplitude of the two-domain cycle keeps decreasing until its other bound also grazes a discontinuity boundary and the cycle disappears completely. 

\section*{Acknowledgements} 
The authors would like to acknowledge the financial support of the European Union (ERC Starting Grant NoVib 307265 and Erasmus+ Programme).
 
\clearpage
\section*{Appendix A: Aerodynamic and structural matrices of the model} 

The matrices appearing in equation~\ref{eq_dof3_qnonlin_freepl} are given by
\begin{equation}
\mathbf{ A}=
\left(
\begin{array}{ccc}
m & S & S_{\beta} \\
S &  I_{\alpha} & I_{\alpha\beta} \\
S_{\beta} & I_{\alpha\beta} & I_{\beta}
\end{array}
\right), \mbox{\hspace{3ex}}
{\bf E}=
\left(
\begin{array}{ccc}
K_h & 0 & 0 \\
0 & K_{\alpha} & 0 \\
0 & 0 & K_{\beta}
\end{array}
\right)
\label{eq_AE}
\end{equation}
\[
\mathbf{ B}=b^2
\left(
\begin{array}{ccc}
\pi & -\pi a b & -T_1 b \\
-\pi a b &  \pi b^2(1/8+a^2) & -(T_7+(c_h-a)T_1)b^2 \\
T_1 b & 2T_{13} b^2 & -T_3 b^2/\pi
\end{array}
\right)
\]
where $a=x_f/b-1$, $b=c/2$, $c_h=x_h/b-1$, $I_{\alpha \beta}=I_{\beta}+b (c_h-a) S_{\beta}$ and the other quantities are given below.
The total aerodynamic damping matrix is given by $\mathbf{D}=\mathbf{D}_1+\Phi(0)\mathbf{ D}_2$ where $\Phi(0)=1-\Psi_1-\Psi_2$, $\Phi(t)=1-\Psi_1e^{-\varepsilon_1Ut/b}-\Psi_2e^{-\varepsilon_2 Ut/b}$ is Wagner's function and
\[
\mathbf{ D}_1=b^2
\left(
\begin{array}{ccc}
0 & \pi  & -T_4  \\
0 & \pi (1/2-a)b &(T_1-T_8-(c_h-a)T_4+T_{11}/2) b  \\
0 & (-2 T_9-T_1+T_4 (a-1/2)) b & b T_{11}/2\pi
\end{array}
\right)
\]
\[
\mathbf{ D}_2=
\left(
\begin{array}{ccc}
2\pi b & 2\pi b^2(1/2-a)  & 2\pi b T_{11}/2\pi  \\
-2 \pi b^2 (a+1/2) & -2 \pi b^3 (a+1/2)(1/2-a) & - b^3 (a+1/2)  T_{11} \\
b^2 T_{12} & b^3 T_{12} (1/2-a) & b^3 T_{12} b T_{11}/2\pi
\end{array}
\right)
\]
The total aerodynamic stiffness is given by $\mathbf{F}=\mathbf{ F}_1+\Phi(0) \mathbf{ F}_2 +\Xi \mathbf{F}_3$ where  $\Xi=\Psi_1\varepsilon_1/b+\Psi_2\varepsilon_2/b$ and
\[
\mathbf{ F}_1=b^2
\left(
\begin{array}{ccc}
0 & 0  & 0  \\
0 & 0 & (T_4+T_{10})  \\
0 & 0  & (T_5-T_4 T_{10})/\pi
\end{array}
\right)
\]
\[
\mathbf{ F}_2=
\left(
\begin{array}{ccc}
0 & 2\pi b  & 2bT_{10}  \\
0 & -2 \pi b^2 (a+1/2) &  -2 b^2 (a+1/2) T_{10}\\
0 & b^2 T_{12}  & b^2 T_{12} T_{10}/ \pi
\end{array}
\right)
\]
\[
\mathbf{F}_3=
\left(
\begin{array}{ccc}
2\pi b & 2\pi b^2 (1/2-a) &  b^2 T_{11}  \\
-2 \pi b^2 (a+1/2) & -2 \pi b^3 (a+1/2)(1/2-a) & -b^3 (a+1/2) T11 \\
b^2 T_{12} & b^3 T_{12} (1/2-a)  & b^3 T_{12} T_{11}/ 2\pi
\end{array}
\right)
\]
The aerodynamic state influence matrix is given by $\mathbf{W}= [2\pi b \mathbf{ W}_0 \mbox{\hspace{2ex}} -2\pi b^2 (a+1/2)\mathbf{ W}_0
 \mbox{\hspace{2ex}} b^2 T_{12}\mathbf{W}_0]^T$ where
 \[
\mathbf{W}_0=
 \left(
 \begin{array}{c}
 -\Psi_1 (\varepsilon_1/b)^2 \\
   -\Psi_2 (\varepsilon_2/b)^2 \\
 \Psi_1 \varepsilon_1 (1-\varepsilon_1 (1/2-a))/b \\
  \Psi_2 \varepsilon_2 (1-\varepsilon_2 (1/2-a))/b \\
  \Psi_1 \varepsilon_1 (T_{10}-\varepsilon_1 T_{11}/2) /\pi b \\
  \Psi_2 \varepsilon_2 (T_{10}-\varepsilon_2 T_{11}/2) /\pi b
 \end{array}
 \right)
 \]
The $T_1$-$T_{14}$ coefficients are defined in Theodorsen~\cite{Theodorsen35} and many other classic aeroelasticity texts. 
\\
\noindent
The structural damping matrix is given by ${\bf D}={{\bf V}^{-1}}^T{\bf B}_{mod}{\bf V}^{-1}$, where ${\bf V}$ are the eigenvectors of the matrix ${\bf A}^{-1}{\bf E}$ and ${\bf B}_{mod}$ is given by
\[
{\bf B}_{mod}=
\left(
\begin{array}{ccc}
2\bar{m}_1 \omega_1 \zeta_1 & 0 & 0 \\
0 & 2\bar{m}_2 \omega_2 \zeta_2 & 0 \\
0 & 0 & 2\bar{m}_3 \omega_3 \zeta_3
\end{array}
\right)
\]
In this latest expression, $\bar{m}_i$ are the diagonal elements of the matrix ${\bf V}^T{\bf A}{\bf V}$, $\omega_i$ are the square roots of the eigenvalues of the matrix ${\bf A}^{-1}{\bf E}$.
\\
Finally, the preload vector $P$ is given by
\[
\mathbf{P}= \left(\begin{array}{ccc}
-2\pi b & 2\pi b^2(a+1/2) &-b^2T_{12}
\end{array}
\right)^T
\]
\clearpage
\section*{Appendix B: Characteristics of the experimental system}  
\begin{table}[h!]
\center
\begin{tabular}{c|cc}
\hline
 & Wing dimensions &  \\
Chord (with flap) & $c$ & 25.4 cm \\
Span & $s$ & 52  cm \\ 
Flexural axis & $x_f$ & 0.25 c  \\
\hline 
& Flap dimensions & \\
Chord (flap alone) & $ - $ & 6.25 cm \\
Span & $s_2$ & 52 cm \\
Hinge axis & $x_h$ & 0.75c \\
\hline
& Inertial parameters & \\
Plunge mass & M & 2.562 kg \\
Pitch inertia & $I_\alpha$ & 0.0181 m kg \\
Control inertia & $I_\beta$ & $2.66 10^{-4}$ m kg \\
Pitch static imbalance & S & 0.0943 m.kg \\
Pitch-Flap inertia product& $I_{\alpha,\beta}$ & 0.0013 mkg\\
Flap static imbalance & $S_\beta$ & 0.0084 mkg \\
\hline
& Stiffness parameters & \\
Plunge stiffness & $K_h$ & 850.7 N/m \\
Pitch stiffness & $K_\alpha$ & 34 Nm/rad \\
Flap stiffness & $K_\beta$ & 1.512 Nm/rad \\
\hline
& Modal parameters & \\
Plunge mode frequency & $f_1$ & 2.9 Hz \\
Pitch mode frequency & $f_2$ & 7.1 Hz\\
Control mode frequency & $f_3$ & 17.0 Hz\\
Plunge mode damping & $\zeta_1$ & 0.87 \% \\
Pitch mode damping & $\zeta_2$ & 1.39\% \\
Flap mode damping & $\zeta_3$ & 0.6 \% \\
\hline
\end{tabular}
\caption{Structural parameters of the experimental system}
\end{table}


\clearpage
\bibliography{refs} 

\end{document}